\documentclass[sigconf]{acmart}
\makeatletter
\def\@ACM@checkaffil{
    \if@ACM@instpresent\else
    \ClassWarningNoLine{\@classname}{No institution present for an affiliation}%
    \fi
    \if@ACM@citypresent\else
    \ClassWarningNoLine{\@classname}{No city present for an affiliation}%
    \fi
    \if@ACM@countrypresent\else
        \ClassWarningNoLine{\@classname}{No country present for an affiliation}%
    \fi
}
\makeatother

\usepackage{fancyhdr}
\usepackage{tablefootnote}
\usepackage{upgreek}
\usepackage{glossaries} 
\usepackage{xcolor}
\usepackage{ifthen}
\usepackage{makecell}
\usepackage{booktabs}
\usepackage{siunitx}

\usepackage{hhline}
\usepackage{float}
\usepackage{algorithm}
\usepackage[us,12hr]{datetime}
\usepackage{soul}
\usepackage{amsmath,amsfonts}
\usepackage{cases}

\usepackage{marginnote} 
\usepackage{graphicx}
\usepackage{textcomp}
\usepackage{todonotes} 
\usepackage{textcomp}
\usepackage[T1]{fontenc}

\usepackage{balance}
\usepackage{flushend}

\usepackage{mathtools}
\usepackage{framed}
\usepackage{setspace}
\usepackage{blindtext}

\usepackage{multicol}
\usepackage{multirow}
\title{Multicols Demo}
\usepackage{tabularx}

\usepackage{enumitem} 
\setlist{nosep}

\usepackage{etoolbox}

\usepackage{fixltx2e}
\usepackage{dblfloatfix}

\AtBeginDocument{%
  \providecommand\BibTeX{{%
    \normalfont B\kern-0.5em{\scshape i\kern-0.25em b}\kern-0.8em\TeX}}}

\usepackage{titlesec}
\titlespacing\section{0pt}{2pt plus 1pt minus 1pt}{3pt plus 1pt minus 2pt}
\titlespacing\subsection{0pt}{2pt plus 1pt minus 1pt}{3pt plus 1pt minus 2pt}
\titlespacing\subsubsection{0pt}{2pt plus 1pt minus 1pt}{3pt plus 1pt minus 2pt}

	\makeatletter
	\g@addto@macro{\normalsize}{%
	  \setlength{\abovedisplayskip}{1pt plus 1pt minus 1pt}
	  \setlength{\belowdisplayskip}{1pt plus 1pt minus 1pt}
	  \setlength{\abovedisplayshortskip}{0pt}
	  \setlength{\belowdisplayshortskip}{0pt}
	  \setlength{\intextsep}{1pt plus 1pt minus 1pt}
	  \setlength{\textfloatsep}{1pt plus 1pt minus 1pt}
	  \setlength{\skip\footins}{4pt plus 1pt minus 1pt}}
	\makeatother

\definecolor{lightyellow}{rgb}{0.980, 0.956, 0.623}
\sethlcolor{lightblue}

\newcommand{\boxbegin} {
	\begin{tcolorbox}[enhanced, frame hidden, colback=gray!50, breakable]
}

\newcommand{\boxend} {
	\end{tcolorbox}
}

\newcommand{\yboxbegin} {
	\begin{tcolorbox}[breakable, enhanced, frame hidden, colback=yellow!50]
}

\newcommand{\yboxend} {
	\end{tcolorbox}
}

\newcommand{\bboxbegin}{
    \begin{mdframed}[style=graybox]
}

\newcommand{\bboxend}{
    \end{mdframed}
}

\newcommand{\yyboxbegin}{
    \begin{mdframed}[style=graybox2]
}

\newcommand{\yyboxend}{
    \end{mdframed}
}

\definecolor{amber}{rgb}{1.0, 0.49, 0.0}
\definecolor{awesome}{rgb}{1.0, 0.13, 0.32}
\definecolor{dollarbill}{rgb}{0.52,0.73,0.4}
\definecolor{moegi}{rgb}{0.357, 0.537, 0.188}
\definecolor{burgundy}{rgb}{0.5, 0.0, 0.13}
\definecolor{ballblue}{rgb}{0.13, 0.67, 0.8}
\definecolor{ups-truck}{rgb}{0.53, 0.28, 0.21}
\definecolor{airforceblue}{rgb}{0.36, 0.54, 0.66}
\definecolor{cadmiumgreen}{rgb}{0.0, 0.42, 0.24}
\definecolor{darkcyan}{rgb}{0.0, 0.55, 0.55}
\definecolor{caribbeangreen}{rgb}{0.0, 0.8, 0.6}
\definecolor{flamingopink}{rgb}{0.99, 0.56, 0.67}
\definecolor{jazzberryjam}{rgb}{0.65, 0.04, 0.37}
\definecolor{mediumpersianblue}{rgb}{0.0, 0.4, 0.65}
\definecolor{coolblack}{rgb}{0.0, 0.18, 0.39}
\definecolor{bleudefrance}{rgb}{0.19, 0.55, 0.91}
\definecolor{ao}{rgb}{0.0, 0.0, 1.0}
\definecolor{babyblueeyes}{rgb}{0.63, 0.79, 0.95}
\definecolor{darkwarmgray}{rgb}{0.2,0,0}
\definecolor{lightblue}{rgb}{0.980, 0.956, 0.623}
\definecolor{joelpink}{rgb}{0.9, 0.0, 0.7}

\newcommand*\circled[1]{\tikz[baseline=(char.base)]{
            \node[shape=circle,draw,inner sep=0pt,fill=black, text=white] (char) {#1};}}
\newcommand*\Rcircled[1]{\tikz[baseline=(char.base)]{
            \node[shape=circle,draw,inner sep=0pt,fill=red, text=white] (char) {#1};}}
\newcommand*\Bcircled[1]{\tikz[baseline=(char.base)]{
            \node[shape=circle,draw,inner sep=0pt,fill=blue, text=white] (char) {#1};}}
\newcommand*\Wcircled[1]{\tikz[baseline=(char.base)]{
            \node[shape=circle,draw,inner sep=0pt,fill=white, text=black] (char) {#1};}}

\newcounter{obs}
\newcounter{take}
\newcounter{obsbox}
\newcounter{chal}
\newcounter{chalbox}
\newif\ifrevision
\revisionfalse

\newif\ifsubmission
\submissionfalse

\newif\ifiscarev
\iscarevfalse

\ifrevision

  \definecolor{moegi}{rgb}{0.357, 0.537, 0.188}

  \newcommand{\ignore}[1]{}

  \newcommand{\mscomment}[1]{}
  \newcommand{\mstodo}[1]{}
  \newcommand{\promotoronecomment}[1]{}
  \newcommand{\promotortwocomment}[1]{}
  \newcommand{\omcomment}[1]{}
  \newcommand{\gs}[1]{} 
\newcommand{\gss}[1]{} 
\newcommand{\gsf}[1]{} 
  
\else

  \ifsubmission   

    \newcommand{\ignore}[1]{}

    \newcommand\mstodo[1]{}
    \newcommand{\mscomment}[1]{}
    \newcommand{\promotoronecomment}[1]{}
    \newcommand{\promotortwocomment}[1]{}
    \newcommand{\omcomment}[1]{}
    \newcommand{\gs}[1]{} 
    \newcommand{\gss}[1]{} 
    \newcommand{\gsf}[1]{} 

  \else           

    \newcommand{\ignore}[1]{}
    
    \newcommand{\mstodo}[1]{\todo[size=\footnotesize,color=jazzberryjam]{\textbf{@TMS:} #1}}
    \newcommand{\mscomment}[1]{\textcolor{amber}{\textbf{[@TMS: {\it#1}]}}}
    \newcommand{\promotoronecomment}[1]{\textcolor{red}{\textbf{[@JSSMW: {\it#1}]}}}
    \newcommand{\promotortwocomment}[1]{\textcolor{red}{\textbf{[@Said Hamdioui: {\it#1}]}}}
    \newcommand{\omcomment}[1]{\textcolor{red}{\textbf{[@OM: {\it#1}]}}}

    \newcommand{\gs}[1]{{\color{red}{#1}}} 
    \newcommand{\gss}[1]{{\color{black}{#1}}} 
    \newcommand{\gsf}[1]{{\color{blue}{GAGAN:#1}}} 

  \fi
\fi

\ifiscarev

    \else
    
\fi

\newcommand\fig[1]{Fig.~{#1}\xspace}

\newcommand\sect[1]{Section~{#1}\xspace}

\newcommand\tab[1]{Table~{#1}\xspace}

\newcommand{\pim}{PIM\xspace}
\newcommand{\pimlong}{Proc\-essing-In-Mem\-ory\xspace}
\newcommand{\cim}{CIM\xspace}
\newcommand{\cimlong}{Comp\-utat\-ion-In-Mem\-ory\xspace}

\newcommand{\vmm}{VMM\xspace}
\newcommand{\vmmlong}{Vec\-tor-Matrix-Mult\-ipli\-cation\xspace}
\newcommand{\mmm}{MMM\xspace}
\newcommand{\mmmlong}{Matrix-Matrix-Multiplication\xspace}
\newcommand{\sram}{SRAM\xspace}

\newcommand{\puma}{PUMA\xspace}
\newcommand{\pumalong}{Programmable Ultra-efficient Memristor-based Accelerator\xspace}

\newcommand{\sotalong}{state-of-the-art\xspace}

\newcommand{\sota}{SotA\xspace}

\newcommand{\ontlong}{Oxford Nanopore Technologies\xspace}
\newcommand{\ont}{ONT\xspace}
\newcommand{\gpulong}{Graphics Processing Unit\xspace}
\newcommand{\gpu}{GPU\xspace}
\newcommand{\fpgalong}{Field-Programmable Gate Array\xspace}
\newcommand{\fpga}{FPGA\xspace}
\newcommand{\cpulong}{Central Processing Unit\xspace}
\newcommand{\cpu}{CPU\xspace}
\newcommand{\hslong}{hardware/software co-design\xspace}
\newcommand{\hs}{HW/SW co-design\xspace}
\newcommand{\mllong}{Machine Learning\xspace}
\newcommand{\ml}{ML\xspace}

\newcommand{\nn}{NN\xspace}
\newcommand{\dnnlong}{Deep Neural Network\xspace}
\newcommand{\dnn}{DNN\xspace}

\newcommand{\cnn}{CNN\xspace}

\newcommand{\lstm}{LSTM\xspace}
\newcommand{\daclong}{digital to analog converter\xspace}
\newcommand{\dac}{DAC\xspace}
\newcommand{\adclong}{analog to digital converter\xspace}
\newcommand{\adc}{ADC\xspace}

\newcommand{\kdlongcapital}{Knowledge Distillation\xspace}
\newcommand{\kdlong}{knowledge distillation\xspace}
\newcommand{\kd}{KD\xspace}
\newcommand{\rsalong}{random sparse adaptation\xspace}
\newcommand{\rsalongcapital}{Random Sparse Adaptation\xspace}
\newcommand{\rsa}{RSA\xspace}
\newcommand{\rvwlongcapital}{Read-Verify-Write\xspace}

\newcommand{\rvw}{R-V-W\xspace}
\newcommand{\vatlong}{variation-aware training\xspace}
\newcommand{\vat}{VAT\xspace}

\newcommand{\oldlong}{open-loop-off-device\xspace}
\newcommand{\old}{OLD\xspace}
\newcommand{\mnist}{MNIST\xspace}

\newcommand{\pcm}{PCM\xspace}

\newcommand{\reram}{ReRAM\xspace}

\newcommand{\swordfish}{Swordfish\xspace}
\newcommand{\accswordfish}{SwordfishAccel\xspace}
\newcommand{\nomitigationhwaccswordfish}{Ide\-al-\-Sword\-fish\-Accel\xspace}
\newcommand{\realhwaccswordfish}{Real\-istic-\-Sword\-fish\-Accel\xspace}
\newcommand{\bonito}{Bonito\xspace}
\newcommand{\gbonito}{Bonito-GPU\xspace}

\newcommand{\guppy}{Guppy\xspace}
\newcommand{\helix}{Helix\xspace}
\newcommand{\dorado}{Dorado\xspace}

\newcommand{\genax}{GenAx\xspace}
\newcommand{\genasm}{GenASM\xspace}
\newcommand{\halcyon}{Halcyon\xspace}
\newcommand{\sacall}{SACall\xspace} 

\newcommand{\partitionandmap}{Partition \& Map\xspace}
\newcommand{\vmmmodelgenerator}{\vmm Model Generator\xspace}
\newcommand{\accuracymitigator}{Accuracy Enhancer\xspace}
\newcommand{\systemevaluator}{System Evaluator\xspace}

\newcommand{\minAccuracyDropAllMitigationsAccswordfishVsBonitoAllErrorsMeasuredSixtyfour}{6.01\%\xspace} 

\newcommand{\writevariationTwentyFive}{25\%\xspace} 
\newcommand{\maxAccuracyDropQuantizationNoSmart}{9\%\xspace}
\newcommand{\accuracyDiffAvgWritevariationTwentyFiveHAEMOvsKLEBS}{0.93\%\xspace} 
\newcommand{\accuracyWriteVariationTenACINETOBACTER}{3.30\%\xspace} 
\newcommand{\accuracyWriteVariationFiftyACINETOBACTER}{87.34\%\xspace} 

\newcommand{\accuracyWriteVariationTenKLEBSIELLA}{3.24\%\xspace}
\newcommand{\accuracyWriteVariationFiftyKLEBSIELLA}{85.76\%\xspace} 

\newcommand{\accuracyDropNoMitigationAccswordfishVsBonitoMin}{18.32\%\xspace}
\newcommand{\accuracyDropNoMitigationAccswordfishVsBonitoMax}{31.32\%\xspace}

\newcommand{\accuracyDropNoMitigationAccswordfishVsBonitoDACsAndDriversACINETOBACTER}{13.32\%\xspace}
\newcommand{\accuracyDropNoMitigationAccswordfishVsBonitoWiresAndsynapticACINETOBACTER}{15.34\%\xspace}

\newcommand{\accuracyDropNoMitigationAccswordfishVsBonitoAddsupACINETOBACTER}{35.96\%\xspace}
\newcommand{\accuracyDropNoMitigationAccswordfishVsBonitoAllApproachTwoACINETOBACTER}{20.32\%\xspace}
\newcommand{\accuracyDropNoMitigationAccswordfishVsBonitoAllApproachOneACINETOBACTERSixtyfour}{20.32\%\xspace}
\newcommand{\accuracyDropNoMitigationAccswordfishVsBonitoAllApproachOneACINETOBACTERTwohundredfiftysix}{26.33\%\xspace}

\newcommand{\accuracyDropNoMitigationAccswordfishVsBonitoCombinedSixtyfourApproachOneMin}{22.19\%\xspace}
\newcommand{\accuracyDropNoMitigationAccswordfishVsBonitoMeasuredSixtyfourApproachOneMin}{24.32\%\xspace}

\newcommand{\throughputImprovementIdealAccswordfishVsBonitoAverage}{413.6$\times$\xspace}

\newcommand{\retraingIterationsQuantizationAverage}{150\xspace}
\newcommand{\accuracyImprovementRetrainingQuantizationAwareAverage}{5\%\xspace}

\newcommand{\accuracyDiffRSAKDandCombinedBelowFifteenPercentWriteVariation}{0.001\%\xspace}

\newcommand{\maxMitigateWriteVariationAfterAllMitigations}{10\%\xspace}

\newcommand{\flashbackAccuracyImprovementVATMitigationAveragedDatasetsSynapticSixtyfour}{6.85\%\xspace}
\newcommand{\flashbackAccuracyImprovementRVWMitigationAveragedDatasetsSynapticSixtyfour}{10.64\%\xspace}
\newcommand{\flashbackAccuracyImprovementRSAKDMitigationAveragedDatasetsSynapticSixtyfour}{10.85\%\xspace}
\newcommand{\flashbackAccuracyImprovementALLMitigationAveragedDatasetsSynapticSixtyfour}{11.84\%\xspace}

\newcommand{\accuracyImprovementVATMitigationAveragedDatasetsDACSixtyfour}{94.22\%\xspace}
\newcommand{\accuracyImprovementRSAKDMitigationAveragedDatasetsDACSixtyfour}{94.32\%\xspace}
\newcommand{\accuracyImprovementVATMitigationAveragedDatasetsSynapticSixtyfour}{87.32\%\xspace}
\newcommand{\accuracyImprovementRSAKDMitigationAveragedDatasetsSynapticSixtyfour}{91.32\%\xspace}

\newcommand{\accuracyImprovementAllMitigationAveragedDatasetsMeasuredTwohundredfiftysix}{22.07\%\xspace}
\newcommand{\accuracyImprovementAllMitigationAveragedDatasetsMeasuredSixtyfour}{16.24\%\xspace}

\newcommand{\defaultPercentageMappedWeightOnChipMemoryRSAKD}{5\%\xspace}

\newcommand{\throughputDropRVWvsGbonitoAveragedDatasets}{30\%\xspace}

\newcommand{\throughputImprovementAccRSAvsGbonitoAveragedDatasets}{5.24$\times$\xspace}
\newcommand{\throughputImprovementAccRSAKDvsGbonitoAveragedDatasets}{25.7$\times$\xspace}

\newcommand{\accuracyRangeAfterRSAKD}{5\%\xspace}

\copyrightyear{2023}
\acmYear{2023}
\setcopyright{acmlicensed}\acmConference[MICRO '23]{56th Annual IEEE/ACM International Symposium on Microarchitecture}{October 28-November 1, 2023}{Toronto, ON, Canada}
\acmBooktitle{56th Annual IEEE/ACM International Symposium on Microarchitecture (MICRO '23), October 28-November 1, 2023, Toronto, ON, Canada}
\acmPrice{15.00}
\acmDOI{10.1145/3613424.3614252}
\acmISBN{979-8-4007-0329-4/23/10}

    \fancypagestyle{firststyle}
    {
        \setlength{\footskip}{40pt}
        \fancyfoot[C]{\thepage}
    }
\begin{document}

\title{Swordfish: A Framework for Evaluating\\Deep Neural Network-based Basecalling\\using Computation-In-Memory with Non-Ideal Memristors}

\author{Taha Shahroodi$^{1}$ \hspace{0.1cm}  Gagandeep Singh$^{2,3}$ \hspace{0.1cm}  Mahdi Zahedi$^1$ \hspace{0.1cm}  Haiyu Mao$^3$ \hspace{0.1cm}  Joel Lindegger$^3$ \hspace{0.1cm} Can Firtina$^3$ \\ \hspace{0.5cm}  Stephan Wong$^1$ \hspace{0.5cm}    Onur Mutlu$^3$ \hspace{0.5cm}  Said Hamdioui$^1$} 
\affiliation{%
  \vspace{0.7em}
  \institution{\textsuperscript{1}TU Delft\quad \textsuperscript{2}AMD Research\quad  \textsuperscript{3}ETH Z{\"u}rich}
}
\email{}


\renewcommand{\shortauthors}{Shahroodi et al.}
\renewcommand{\shorttitle}{\swordfish}

\renewcommand{\authors}{Taha Shahroodi, 
Gagandeep Singh,
Mahdi Zahedi,
Haiyu Mao,
Joel Lindegger,
Can Firtina,
Stephan Wong,
Onur Mutlu,
Said Hamdioui}

\begin{abstract}

\emph{Basecalling}, an essential step in many genome analysis studies, relies on large \dnnlong{}s (\dnn{}s) to achieve high accuracy. Unfortunately, these \dnn{}s are computationally slow and inefficient, leading to considerable delays and resource constraints in the sequence analysis process. A \cimlong (\cim) architecture using memristors can significantly accelerate the performance of \dnn{}s. However, inherent device non-idealities and architectural limitations of such designs can greatly degrade the basecalling accuracy, which is critical for accurate genome analysis. To facilitate the adoption of memristor-based \cim designs for basecalling, it is important to (1) conduct a comprehensive analysis of potential \cim architectures and (2) develop effective strategies for mitigating the possible adverse effects of inherent device non-idealities and architectural limitations. 

This paper proposes \swordfish, a novel hardware/software co-design framework that can effectively address the two aforementioned issues. \swordfish incorporates seven circuit and device restrictions or non-idealities from characterized real memristor-based chips. \swordfish leverages various \hslong solutions to mitigate the basecalling accuracy loss due to such non-idealities. To demonstrate the effectiveness of \swordfish, we take \bonito, the \sotalong (i.e., accurate and fast), open-source basecaller as a case study. Our experimental results using \swordfish show that a \cim architecture can realistically accelerate \bonito for a wide range of real datasets by an average of \throughputImprovementAccRSAKDvsGbonitoAveragedDatasets, with an accuracy loss of \minAccuracyDropAllMitigationsAccswordfishVsBonitoAllErrorsMeasuredSixtyfour.

\end{abstract}


\maketitle
\thispagestyle{firstpage}

\section{Introduction} \label{sec:introduction}

\emph{Basecalling} is the first computational step required to translate noisy electrical signals generated by modern sequencing machines to strings of DNA nucleotide bases (i.e., \{A, C, G, T\}), also known as DNA reads or simply reads~\cite{van2018third, ardui2018single, jain2018nanopore,  kchouk2017generations, weirather2017comprehensive, wang2021nanopore, pages2022comprehensive, senol2019nanopore, ALSER20224579-MohammedSurvey-LongvsShortReadsPipelines}. The accuracy of basecalling directly affects the overall accuracy and the computational effort (in terms of required algorithms and their complexity and runtimes) of subsequent genome analysis steps. The speed of basecalling also determines how fast one can run through all computational steps of a genomic study~\cite{wick2019performance-Guppy, senol2019nanopore, singh2022framework-RUBICONQABASSkipClipRUBICALL-gaganbasecaller}. Therefore, accurate and fast basecalling is critical for advancing genomic studies that hold the key to unlocking the potential of precision medicine, facilitating virus surveillance, and driving advancements in healthcare and science~\cite{ginsburg2018precision, aryan2020moving, clark2019diagnosis, kingsmore2022genome, ginsburg2009genomic, bloom2021massively, quick2016real, yelagandula2021multiplexed, le2013selected, nikolayevskyy2016whole, wooley2010primer, alkan2009personalized-PrecisionMedicine, ashley2016towards-PrecisionMedicine,chin2011cancer-PrecisionMedicine, ellegren2014genome-PrecisionMedicine, alvarez2017next-ForensicSciences, ALSER20224579-MohammedSurvey-LongvsShortReadsPipelines, mutlu2023accelerating-DAC2023}.

Current \sotalong (\sota) basecallers leverage \dnnlong{}s (\dnn{}s) to achieve high accuracy~\cite{dias2019artificial, zhang2020nanopore, rang2018squiggle, bonito2020, xu2021fast-FastBonito, singh2022framework-RUBICONQABASSkipClipRUBICALL-gaganbasecaller}. However,  \sota \dnn-based basecallers encounter different shortcomings when implemented using different approaches. Specifically, \dnn-based basecaller designs on \cpulong{}s (\cpu{}s) and \gpulong{}s (\gpu{}s) face multiple major shortcomings: (1)~they are computationally intensive and slow~\cite{wick2019performance-Guppy, senol2019nanopore, singh2022framework-RUBICONQABASSkipClipRUBICALL-gaganbasecaller}, (2)~they require extensive data movement between the processor and memory~\cite{liu2018processing, boroumand2018google, Google_Neural_Network_Models_for_Edge_Devices-PACT2021}, and (3)~they are limited by the use of costly hardware, such as expensive \sram memories that require 6 transistors for storing only 1 bit of information~\cite{courbariaux2015binaryconnect-binaryconnect-MahdiBCIM27, qin2020binary-BNNSurveyBNNinEmbedded-MahdiBCIM5}. When implemented on a hardware accelerator, these \dnn{}-based basecallers face two other limitations: (1)~They rely on costly floating-point (FP) computations, which place high demands on the required system's memory bandwidth and compute units with FP capability. This makes hardware acceleration difficult due to the large number and size of neural network model parameters. (2)~They use costly \mllong (\ml) techniques such as skip connections\footnote{Skip connection is an \ml technique that allows skipping a few neural network layers and forwarding the output to the input of a layer further ahead.}~\cite{SkipConnection-Inceptionv4_Inception_ResNet-szegedy2017inception,  bonito2020, xu2021fast-FastBonito}, leading to added computation, memory, and storage overheads (e.g., to store the activation parameters that are fed to the last layers of the \nn)~\cite{singh2022framework-RUBICONQABASSkipClipRUBICALL-gaganbasecaller}. Therefore, over the past decade, both industry and academia~\cite{shafiee2016isaac, chi2016prime, lee2009architecting, shahroodi2023ACaseforGenomeAnalysisWhereGenomesResides, singh2022cim-sttmramencryptionBNNCIMAbhairaj, prezioso2015training} have explored the use of \cimlong (\cim)\footnote{Interchangeably, also referred to as \pimlong (\pim)~\cite{mutlu2023modern}.} using memristor-based devices to accelerate \dnn{}s. 

This growing interest in using \cim for resolving the shortcomings of \dnn{}s is driven by two main factors: (1)~the potential of the \cim paradigm to process data where it resides to reduce the large performance and energy overheads of data movement and (2)~the analog operational properties of these nanoscale emerging technologies (e.g., memristors) that intrinsically support efficient \vmmlong (\vmm), multiple of which are used to implement a \mmmlong (\mmm) that is the most dominant operation in \dnn{}s. However, the memristor-based \cim solutions for basecalling can greatly degrade the \dnn inference accuracy due to (1)~the limited quantization levels supported by memristor devices~\cite{shafiee2016isaac, chi2016prime} and (2)~non-idealities of memristive devices and circuits used to adopt memristor-based memory arrays, such as sneak paths~\cite{shi2020research-AbhairajISCAS47-Sneakpath1, hu2018memristor-AbhairajISCAS32-Sneakpath2} and the non-linearity of peripheral circuitry~\cite{zhang2019design-AbhairajISCAS42-Nonlinearity1, karpov2007nucleation-AbhairajISCAS43-Nonlinearity2, 8935518moinuddin2019low-AbhairajISCAS44-Nonlinearity3}. To propose viable solutions for accelerating the large-scale \dnn{}-based basecallers, these aspects must be considered at all computing stack layers, i.e., application, architecture, and device.  Such considerations are only possible with a framework capable of evaluating the impact of the non-idealities in memristor-based \cim architecture on the end-to-end basecalling accuracy. This framework should also be able to account for the overhead that the solutions to overcome the accuracy loss may bring.

To this end, we propose \emph{\swordfish}, a modular and extensible \hslong framework that allows us to (1)~evaluate the impact of memristor non-idealities and \cim limitations on the accuracy and performance of basecalling and (2)~investigate potential mitigation techniques and measure their effect on accuracy for each non-ideality (\textbf{Contribution \#1}). \swordfish is used to investigate the acceleration of basecalling via emerging computing paradigms and technologies. Specifically, with \swordfish, we comprehensively investigate the potential of accurate acceleration of a \sota basecaller (\bonito) on a \sota \cim architecture (\puma~\cite{ankit2019puma}) by accounting for the non-idealities of the underlying devices and technologies of the underlying architecture, for the first time (\textbf{Contribution \#2}). \swordfish integrates real-world applications with multiple critical comparison metrics, distinct mitigation strategies to tackle the challenges of novel hardware, and comprehensive real measurements to guide the modeling of memristors. Our evaluations using \swordfish show that on a wide range of real genome datasets, \puma accelerates \bonito, a \sota basecaller, by an average of \throughputImprovementAccRSAKDvsGbonitoAveragedDatasets realistically (i.e., the average throughput improvement is \throughputImprovementAccRSAKDvsGbonitoAveragedDatasets when we consider essential mitigation techniques to prevent huge accuracy loss). This performance still comes at the cost of a \minAccuracyDropAllMitigationsAccswordfishVsBonitoAllErrorsMeasuredSixtyfour accuracy loss (\sect{\ref{sec:evaluations_notSmart_and_smartMitigations}}). Our evaluations also yield several key suggestions and recommendations for \dnn, hardware, and system designers of future emerging accelerators with memristors for \dnn-based basecallers and other applications that have two most important metrics (e.g., accuracy and performance) to consider in their evaluation (\textbf{Contribution \#3}). Specifically, our investigation using \swordfish results in multiple unique insights: (1)~Our results challenge the prevalent assumption that \dnn-based applications will automatically succeed on memristor-based \cim due to inherent redundancy in large neural networks, (2)~combining mitigation techniques at only one abstraction level (e.g., circuit or system level) does not necessarily improve the accuracy loss as they can potentially go against each other, and (3) combining multiple mitigation techniques at the circuit and system levels can offset the accuracy loss induced by non-idealities significantly.  

\section{Background and Motivation} \label{sec:background_and_motivation}

This section briefly discusses the necessary background and motivation for this work. We refer the reader to comprehensive reviews~\cite{mutlu2023modern, pages2022comprehensive, branton2008potential, hamdioui2015memristor, ALSER20224579-MohammedSurvey-LongvsShortReadsPipelines} for more details.

\subsection{Genome Sequencing Pipeline}
\label{subsec:genome_sequencing_pipeline-Background}

The genome sequencing pipeline consists of computational steps we employ to acquire genome sequences as strings of DNA characters (i.e., \{A, C, G, T\})~\cite{van2018third, ardui2018single, jain2018nanopore,  kchouk2017generations, weirather2017comprehensive, wang2021nanopore, pages2022comprehensive, senol2019nanopore, ALSER20224579-MohammedSurvey-LongvsShortReadsPipelines} for subsequent analysis in bioinformatics, e.g., cell type identification, identification of marker genes, and variant detection.

Although, currently, the most available data and tools in the genomics realm are for short reads~\cite{fujiki2018genax, cali2020genasm} (mainly produced by Illumina sequencers), working with highly accurate long genome sequences is generally favorable as they reduce the computational cost of reconstructing the genome. For this reason, there is a large momentum towards accurate long-read sequencing~\cite{ALSER20224579-MohammedSurvey-LongvsShortReadsPipelines}. Our work focuses on finding solutions and analysis tools that target long reads while also not discarding tools (e.g., \genax~\cite{fujiki2018genax} and \genasm~\cite{cali2020genasm}),  designed for short reads. A leading method for long-read sequencing is the nanopore sequencing technology. Nanopore sequencers \cite{gridIon, PromethION, MinION} translate raw signal squiggles into bases (A, C, G, T) using complex neural networks. Today, \ontlong (\ont) is a company that produces the most commonly used sequencers based on Nanopore technology. 

\begin{figure}[htbp]
\vspace{0.5em}
\centering
    \includegraphics[width=1\linewidth]{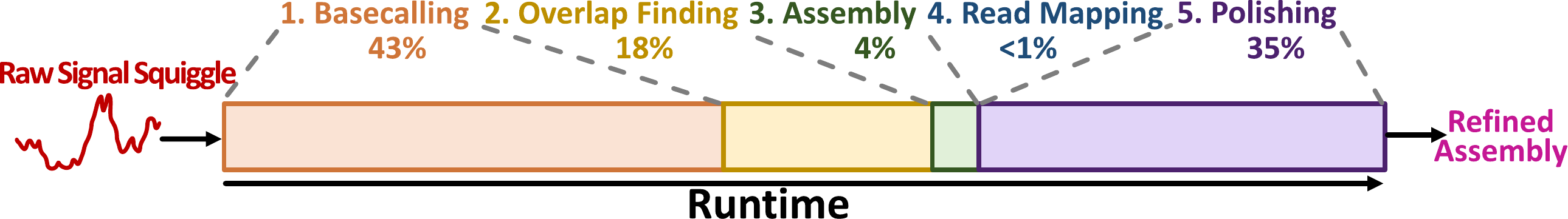}
    \caption{Overview of the nanopore genome sequencing pipeline and execution time breakdown of different steps.}    \label{fig:nanopore_genome_sequencing_pipeline_and_executionTime_breakdown-background_and_motivation}
\end{figure}

\fig{\ref{fig:nanopore_genome_sequencing_pipeline_and_executionTime_breakdown-background_and_motivation}} illustrates the nanopore genome sequencing pipeline~\cite{senol2019nanopore} and the placement and execution time breakdown of each of its steps. We use \sota tools for each step and run the tool on the datasets described in \sect{\ref{sec:experimental_setup_and_methodology}}.

We make two main observations. First, basecalling is the first computational step in the pipeline. Second, basecalling dominates the execution time of a single run in the pipeline. These steps make up more than 40\% of the entire execution time. Our empirical observation aligns with those in prior works~\cite{dunn2021squigglefilter, senol2019nanopore, lou2020helix}. 

\subsection{Basecalling}
\label{subsec:basecalling-Background}

Basecalling is responsible for converting raw electrical signals produced by a nanopore sequencer to digital genome symbols, i.e., [A, C, G, T]~\cite{van2018third, ardui2018single, jain2018nanopore, kchouk2017generations}. Recent works~\cite{flappie2018, Scrappie2019, wick2019performance-Guppy, bonito2020} heavily investigate the use of \dnn{}s for basecalling as they can provide high accuracy than Hidden Markov Model (HMM) based techniques~\cite{Metrichor2017}.

There are generally two approaches for improving the accuracy and/or performance of a basecaller: 1)~software-based and 2)~hardware-based. Software-based methods propose new algorithms (e.g., \dnn{}s~\cite{bonito2020, xu2021fast-FastBonito, Scrappie2019} instead of HMMs~\cite{Metrichor2017}) or faster and/or smaller \dnn architectures~\cite{xu2021fast-FastBonito, singh2022framework-RUBICONQABASSkipClipRUBICALL-gaganbasecaller}. Hardware-based approaches propose various hardware platforms for the target algorithm (i.e., \dnn or HMM) to improve performance with (hopefully) small impact on accuracy~\cite{lou2020helix, singh2022framework-RUBICONQABASSkipClipRUBICALL-gaganbasecaller}.

We observe four main shortcomings in \sota basecallers, which limit their execution time and/or hardware acceleration:

\begin{itemize}[leftmargin=*]
    \item \sota basecallers are slow and energy inefficient. For example, Guppy basecalls 3 Giga basepairs (Gbps) in $\sim$6 hours while a following step in the genomics pipeline, such as read mapping using minimap2~\cite{li2016minimap-minimapandMiniasm} takes only $\sim$0.11 hours~\cite{singh2022framework-RUBICONQABASSkipClipRUBICALL-gaganbasecaller}.

    \item \sota basecallers use \dnn models with costly skip connections \cite{SkipConnection-Inceptionv4_Inception_ResNet-szegedy2017inception}. For example, \bonito needs an additional $\sim$21\% of model parameters (along with associated memory and storage overheads) for skip connections and requires additional computation on them. Note that a skip connection permits bypassing certain layers within the neural network, transmitting the output of one layer as the input to subsequent layers~\cite{SkipConnection-Inceptionv4_Inception_ResNet-szegedy2017inception}. These connections are costly because they (1) typically force the network to perform additional computation, for example, to match the channel sizes, (2) incur extra memory and storage overhead, as they require storing the activation parameters that are fed to the later layers~\cite{boroumand2018google, Google_Neural_Network_Models_for_Edge_Devices-PACT2021}, and (3) incur additional off-chip data movement overhead when these networks are run on conventional processor-centric hardware platforms, like \cpu{}s and \gpu{}s.

    \item \sota basecallers exploit 32-bit floating point precision for their model parameters~\cite{bonito2020, xu2021fast-FastBonito, wick2019performance-Guppy}. This effectively increases (1) the required bandwidth and processing units, e.g., with FP compute capability, and (2) inefficiency in the hardware realization of the underlying models.

    \item \sota basecallers incur expensive data movement between the computation units and the memory units~\cite{liu2018processing, lou2020helix, singh2022framework-RUBICONQABASSkipClipRUBICALL-gaganbasecaller}.

\end{itemize}

We emphasize that 40\% of execution time spent on basecalling (\sect{\ref{subsec:genome_sequencing_pipeline-Background}}), the first and arguably most critical step in the pipeline, is significant and worth accelerating. Today's best basecallers often underperform on \sota systems, generating bottlenecks. A potentially 40\% decrease in genome analysis runtime implies a proportional reduction in power and energy, which is critical considering the extensive data and computational demands of modern genome analysis systems. Therefore, optimizing basecalling contributes greatly to improving the efficiency and sustainability of the genomics pipeline.

\subsection{Memristor-based \cim and Associated Non-Idealities} 
\label{subsec:memristors_noises-Background}

Resistive memories or memristive devices, such as ReRAM, PCM, and STT-MRAM~\cite{waser2009redox, lee2010phase, HDC-CIM-IBM, singh2022cim-sttmramencryptionBNNCIMAbhairaj}, have recently been introduced as suitable candidates for both storage and computation units that can efficiently perform vector-matrix multiplication~\cite{xia2019memristive} and logical bulk bit-wise operations~\cite{xie2017scouting, cheng2019functional, shahroodi2023SieveMem-tahamichaelshahroodiSieveMemASAP2023, li2016pinatubo, shahroodi2023RattlesnakeJake}, as they can follow Kirchhoff’s law inherently~\cite{strukov2008missing-HP}. Therefore, many recent works~\cite{shafiee2016isaac, xie2017scouting, ankit2019puma, chi2016prime, cheng2019functional, zahedi2022system-MahdiVLSISOCSpecialSession, shahroodi2023Lightspeed-BNNOPCMLBRDate2023, SparseMem_DATE2023_MahdiZahedi, Mahdi_arithmetic_MMM_twosComplement_IEEEAccess_2022} exploit these devices in their \cim architectures. Memristor devices also enjoy non-volatility, high-density, and near-zero standby power~\cite{xie2017scouting, li2016pinatubo, singh2022cim-sttmramencryptionBNNCIMAbhairaj}.

\begin{figure}[htbp]
\centering
    \includegraphics[width=1\linewidth]{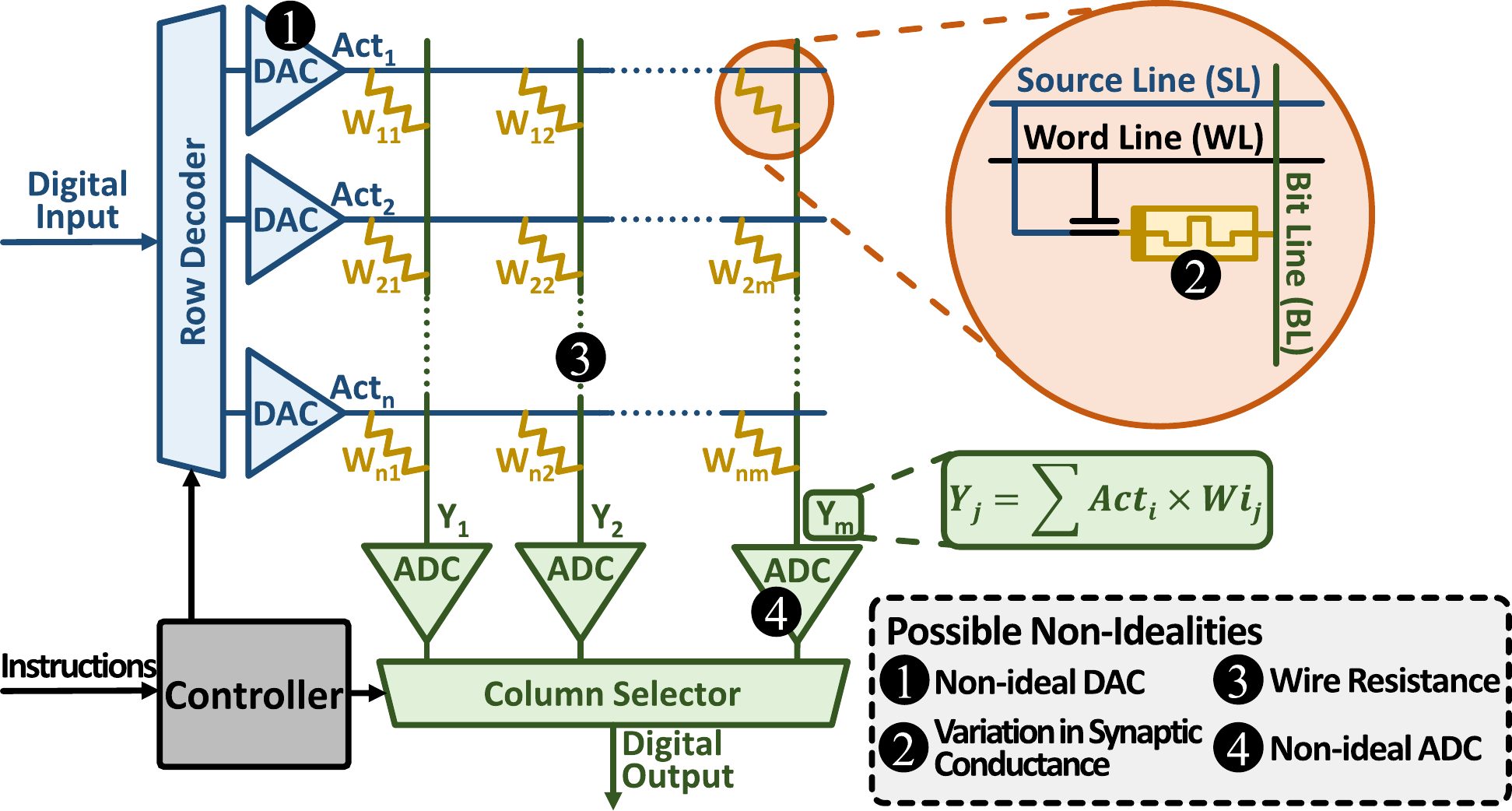}
    \caption{Overview of memristor-based crossbar arrays and possible non-idealities.}
    \label{fig:VMM_using_memristors_crossbar_array_with_nonidealities-background_and_motivation}
\end{figure}

A typical memristor-based memory crossbar capable of \vmm and other logical operations is shown in \fig{\ref{fig:VMM_using_memristors_crossbar_array_with_nonidealities-background_and_motivation}}~\cite{shafiee2016isaac, ankit2019puma, chi2016prime, xie2017scouting, cheng2019functional} alongside its possible non-idealities.

This memristor-based structure can suffer from at least four types of non-idealities or variations that can eventually affect the results of the enabled \vmm operation, i.e., lead to errors in the \vmm result: (1) The non-ideal \daclong{} (\dac{}), due to the effective resistive load (known as $R_{Load}$) in its circuit~\cite{jain2020rxnn-NonIdealities-ADCDACVariations}, (2) Variation of synaptic conductance, which includes both imperfect programming operation (commonly known as write variations) and the process variation that exist in memristors~\cite{alibart2012high-DevicetoDeviceVatriation1, chen2014rram-DevicetoDeviceVatriation2, lee2019exploring-DevicetoDeviceVatriation3, zhang2011stt-DevicetoDeviceVatriation4}, (3) The wire resistance and sneak paths, due to imperfect wires (i.e., wires with different resistances) and the changes in the voltages of the internal nodes while performing a \vmm operation~\cite{jeong2017parasitic-wireParasitics-SumitAnteneh, zhang2011stt-DevicetoDeviceVatriation4}, and (4) non-ideal sensing circuit or \adclong{}s (\adc{}s), due to rigid or hard-to-accurately-change references used for distinguishing/sensing the end result~\cite{jain2020rxnn-NonIdealities-ADCDACVariations, zahedi2022system-MahdiVLSISOCSpecialSession}. Our work focuses on these specific non-idealities inherent to memristor technologies in a \cim architecture. While we do not explicitly address other circuit challenges and non-idealities, we acknowledge their presence and the existing solutions developed to mitigate them in electronic systems. For example, crosstalk~\cite{Crosstalk_in_VLSI_interconnections, Crosstalk_reduction_for_VLSI, Measurement_of_multipath_interference_in_the_coherent_crosstalk_regime}, which involves interference between adjacent circuit traces or wires, can indeed lead to data corruption and compromise information integrity. However, we focus on the specific non-idealities relevant to our hardware architecture, not crosstalk. Note that industry-standard techniques, such as shielding and layout design, decoupling components, ground and power distribution, signal timing and margins, ECC and scrubbing, isolation and shielding, and crosstalk-aware clock distribution, have been extensively studied and developed to mitigate crosstalk issues. We assume that similar techniques can be applied to address any potential crosstalk concerns in memristor-based \cim systems.

Recent works~\cite{shafiee2016isaac, burr2017neuromorphic-VMMinDNN1, chi2016prime, ni2017energy-VmminDNN2, ankit2019puma} report impressive performance and energy improvements for \dnn models executed on memristor-based \cim architectures, mainly assuming idealized underlying hardware. Moreover, \dnn{}s are known to be resilient to some noise~\cite{gupta2015deep-DNNErrorResiliency1, venkataramani2014axnn-DNNErrorResiliency2, tann2017hardware-DNNErrorResiliency3, koppula2019eden, han2015deepcompression, ueyoshi2018quest}. However, since memristor-based \cim architectures are indeed non-ideal and the resiliency of \dnn{}s has a limit, to decide whether or not these platforms are indeed suitable for realizing our \dnn-based basecaller, one needs to evaluate the impact of these non-idealities on the end-to-end application accuracy and account for the overhead that the solutions to overcome the accuracy loss may bring. Such a framework is missing among prior works and is a contribution of our work (\sect{\ref{sec:swordfish_Framework}}). 

\subsection{Programmable Inference Architecture}
\label{subsec:programmable_accelerator_architectures_memristorbased_PUMA_ISAAC-Background}

\puma (\pumalong)~\cite{ankit2019puma, pumafunctionalsimulator2020, pumacompiler2019} is a complete set of (micro)architecture, simulator, and compiler that supports the execution of many \ml applications, using memristor crossbars enhanced with general-purpose execution units. \puma uses a spatial architecture and provides the necessary programmability and generality to execute a wide range of \ml-based applications on memristor-based crossbars. For evaluations in \swordfish, we assume an \puma-based architecture for two reasons. First, \puma supports all the necessary types of \nn layers in basecallers: CNN, LSTM, and linear. This is especially handy for our main target basecaller, \bonito. Second, the architecture, simulator, and compiler are open-sourced~\cite{pumafunctionalsimulator2020, pumacompiler2019} and well-documented for an extension, unlike many other rich architectures.

\section{Swordfish Framework} \label{sec:swordfish_Framework}

\swordfish is a framework designed to guide the evaluation of \cim designs for \dnn-based basecallers.

\subsection{Swordfish Overview} 
\label{subsec:overview-swordfish_Framework}

\fig{\ref{fig:mechanism_framework_overview_vertical-swordfish_Framework}} presents an overview of the \swordfish framework. \swordfish consists of 4 key modules:

\begin{itemize} [leftmargin=*]
    \item \circled{1} \emph{\partitionandmap} module that partitions and maps the \vmmlong (\vmm) operations of the target \dnn-based basecaller to the underlying \cim platform, 
    \item \circled{2} \emph{\vmmmodelgenerator} module that generates an end-to-end model for possible non-idealities and errors of a \vmm operation considering the underlying technology in the \cim design, 
    \item \circled{3} \emph{\accuracymitigator} module that implements online and offline mitigation techniques to counter accuracy loss, and 
    \item \circled{4} \emph{\systemevaluator} module that analyzes the accuracy and throughput of basecaller while also providing an area overhead. 
\end{itemize}

\begin{figure}[htbp]
\vspace{0.5em}
\centering
    \includegraphics[width=1\linewidth]{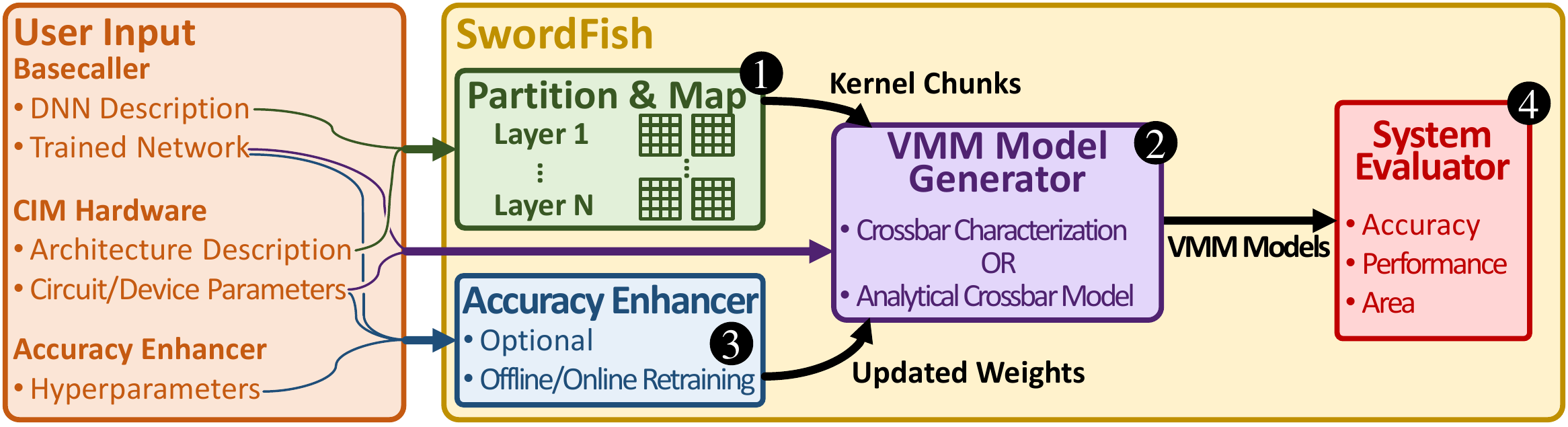}
    \caption{Overview of \swordfish framework.}
    \label{fig:mechanism_framework_overview_vertical-swordfish_Framework}
\end{figure}

We emphasize that the accuracy analysis in the \systemevaluator module is critical and unlike evaluations of conventional platforms, e.g., \fpgalong{}s (\fpga{}s) or \gpu{}s. Its importance stems from the abundance of the underlying non-idealities, variations, limitations, and hardware perturbations of the emerging hardware paradigms~\cite{jiang2020device}. From now on, we refer to the proposed framework as \emph{\swordfish} and the actual implemented memristor-based \cim design for our target basecaller \bonito as \emph{\accswordfish}.

\subsection{\partitionandmap} 
\label{subsec:partition_and_map}

To run the \dnn of a basecaller on a \cim architecture, one should map each of the \vmm operations in the target \dnn to the analog memory arrays and the rest of the operations to the digital peripheral circuitry. The \partitionandmap module takes care of this task in \swordfish by dividing individual functions of the basecaller into the analog or digital components of the underlying architecture. This process is required one time for every basecaller and has two steps.

In the first step, \swordfish decides which memory crossbars will perform each \vmm operation of each layer. For \bonito basecaller, \swordfish decides which memory crossbars handle the \vmm of the first convolutional layer and which crossbars are responsible for the \vmm{}s of the following \lstm and linear layers. \swordfish assumes that all the underlying crossbars have the same size and readout peripheral circuitry  (e.g., \adc{}s).

In the second step, \swordfish decides how it maps the weights to each crossbar. \swordfish supports different programming/writing techniques for memristor devices, such as write-read-verify (WRV) and Set/Reset pulse programming.

In mapping and evaluation, \swordfish makes the following widely common design choices:

\begin{itemize}[leftmargin=*]
    \item The input streams into the first layer of \dnn. \swordfish does not divide the input into chunks and leaves this task to the host. Doing so helps \swordfish to evaluate the maximum throughput of a basecaller~\cite{bonito2020, lou2020helix}, independently of the input size.

    \item The next layer starts its computation as soon as the previous layer of the basecaller produces enough values. This is also a common assumption for evaluating the maximum possible throughput of a \dnn in simulation~\cite{ankit2019puma, shafiee2016isaac}.

    \item Multiple crossbar arrays can be simultaneously active and perform the necessary operations (\vmm and other operations necessary for the target \dnn, such as activation. This assumption ensures that full chip utilization is not limited due to power constraints. One can consider this parallelism to be analogous to the concurrent activation of multiple subarrays in different banks and bank groups in traditional DRAM~\cite{LISA-HPCA2016-inter_subarray_datamovement, RowClone-MICRO2013-in_DRAM_bulk_data_copy, seshadri2017ambit}.

    \item \swordfish optimizes its design decisions for the highest achievable accuracy, throughput, and memory utilization in the stated order. This is a common priority order for optimizations in basecallers~\cite{bonito2020, dorado2022, lou2020helix}.

\end{itemize}

\subsection{\vmmmodelgenerator} \label{subsec:vmm_model_generator}

\begin{figure*}[!b]
\centering
\setcounter{figure}{4}
    \includegraphics[width=1\linewidth]{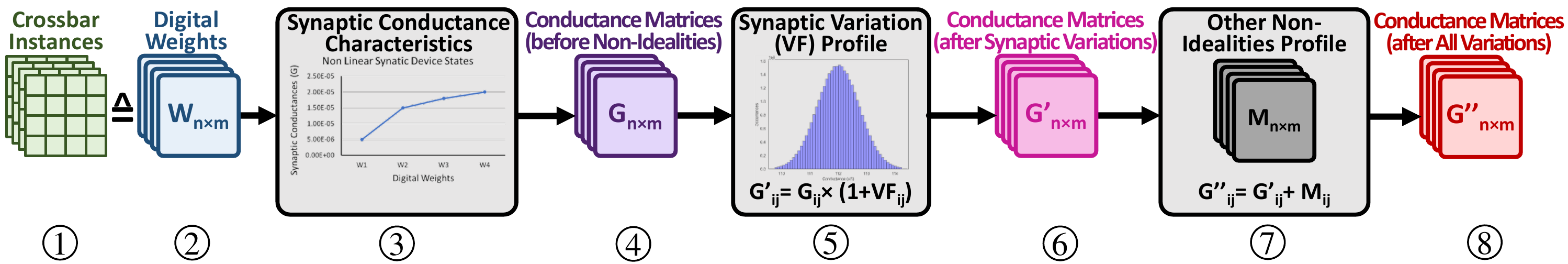}
        \caption{An overview of modeling crossbar non-idealities in \swordfish.}
    \label{fig:crossbar_nonidealities_model_generator-FCM_nonideal_memory_crossbar_model-modeling_constraints_and_nonidealities_error_sources}
\end{figure*}

\vmmmodelgenerator is responsible for generating the non-ideal output per each \vmm required by the basecaller. \vmmmodelgenerator differentiates between constraints and non-idealities. This is essential in a \cim design where non-idealities or constraints do not necessarily lead to a loss in the accuracy of the application. To model the effect of these constraints and non-idealities on the accuracy of an application, \swordfish considers them at the lowest-level building block where they aggregate, i.e., where their results merge. In a memristor-based \cim architecture for a \dnn-based basecaller, such an effective place to consider the effects of constraints and non-idealities is the \vmm operation output. Therefore, the \vmmmodelgenerator in \swordfish focuses on assessing the effects of each factor on a \vmm operation, while our evaluations and analyses assess the end-to-end basecalling metric.

This module takes three types of inputs. First, it takes the results of the previous module (i.e., \circled{1} \partitionandmap in \fig{\ref{fig:mechanism_framework_overview_vertical-swordfish_Framework}}) to determine the size of the \vmm. Second, it takes the circuit and device description (i.e., constraints and non-idealities) that can affect accuracy. Examples inputs in this category are (1) the level of quantization, (2) the circuit variations (e.g., in inputs (e.g., \dac{}s), wires, and outputs (e.g., \adc{}s) device), and (3) device variations. Third, it takes the weights of the target basecaller, which can be provided directly by the user or the \accuracymitigator module that applies multiple training mechanisms (\sect{\ref{subsec:accuracy_mitigator}}). The module outputs the non-ideal output vector per each input vector and weight matrix (i.e., the expected vector result for a \vmm).

\swordfish supports two different approaches for modeling a \vmm. The first approach is to use a pre-calculated library of measurements on actual devices. The second approach is to use an analytical model (e.g., a fast crossbar model (FCM)~\cite{jain2020rxnn-NonIdealities-ADCDACVariations}). \sect{\ref{sec:evaluations_notSmart_and_smartMitigations}} evaluates these approaches separately.

In the first approach, \swordfish queries a library that, for a given array size and input vector, returns an output vector randomly chosen from many ($\ge10^4$) possible outputs based on measurements on an actual crossbar with the same dimensions as the length of the active input vector. The measurements in the library already contain all the possible non-idealities in the target \vmm operation, i.e., non-idealities that may arise from \dac{}s, \adc{}s, circuits, and devices in the crossbar. One can build this library by measuring multiple tiles several times. For each of these measurements, one should program the initial values of memristors within a tile with the weight values of the target \dnn to be evaluated on \swordfish. In this paper, the distinct initial resistance states are based on the \bonito basecaller~\cite{bonito2020}. The random choice from the library aims to account for variations and non-idealities among different memristor-based tiles, which can arise from different initial values of each memristor device and/or manufacturing differences. By integrating real measurements and accounting for tile-to-tile differences, we believe our methods accurately reflect non-ideality distribution in practical settings. Although this approach accurately represents the \vmm operation considering many possible non-idealities, it lacks the flexibility of separately studying or measuring the effects of each possible error due to different non-idealities. This approach is also limited to the crossbar configurations (for example, crossbars of 64$\times$64 and 256$\times$256) to whose measurements one has access (\sect{\ref{sec:experimental_setup_and_methodology}}).

In the second approach, \swordfish utilizes existing analytical models that are available for \adc{}s, \dac{}s, and variation profiles of the underlying devices in the crossbar. \fig{\ref{fig:nonIdealities_DAC_Crossbar_ADC_FCMOvervies-modeling_non-idealities-modeling_constraints_and_nonidealities_error_sources}} illustrates the steps \swordfish uses in its \vmmmodelgenerator for this approach.

\begin{figure}[htbp]
\centering
\setcounter{figure}{3}
    \includegraphics[width=1\linewidth]{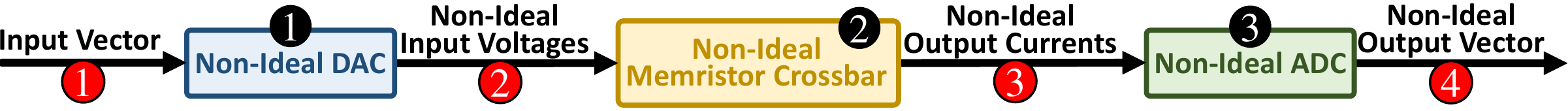}
    \caption{An overview of the \vmmmodelgenerator's second approach: using analytical models.} 
    \label{fig:nonIdealities_DAC_Crossbar_ADC_FCMOvervies-modeling_non-idealities-modeling_constraints_and_nonidealities_error_sources}
\end{figure}

In \fig{\ref{fig:nonIdealities_DAC_Crossbar_ADC_FCMOvervies-modeling_non-idealities-modeling_constraints_and_nonidealities_error_sources}}, \swordfish applies the analytical model for a non-ideal \dac model (\circled{1}) to the input vector of the \vmm operation (\Rcircled{1}) and obtains the non-ideal input voltages as the output vector (\Rcircled{2}). \swordfish then applies this new vector to a crossbar with an updated non-ideal weight matrix (\circled{2}), where non-idealities have been applied to the original weight matrix (from the \vmm operation) based on the expected variations of each cell, which are usually obtained based on generic characterization of memristor-based crossbar arrays, i.e., without any peripheral circuitry or target weights specific to a particular \dnn. The output is considered a non-ideal output current (\Rcircled{3}) that \swordfish applies to a model of non-ideal \adc (\circled{3}) and obtains the output vector (\Rcircled{4}), an output vector that might contain some errors.

\fig{\ref{fig:crossbar_nonidealities_model_generator-FCM_nonideal_memory_crossbar_model-modeling_constraints_and_nonidealities_error_sources}} presents an overview of how \swordfish models the crossbar non-idealities for the second approach (i.e., the analytical model in the \vmmmodelgenerator module) (\circled{2} in \fig{\ref{fig:nonIdealities_DAC_Crossbar_ADC_FCMOvervies-modeling_non-idealities-modeling_constraints_and_nonidealities_error_sources}}). For this, \swordfish first takes the crossbar instances (\Wcircled{1} in \fig{\ref{fig:crossbar_nonidealities_model_generator-FCM_nonideal_memory_crossbar_model-modeling_constraints_and_nonidealities_error_sources}}) from the \partitionandmap module. \swordfish considers these crossbar instances as separate matrices with digital weights (\Wcircled{2}). Then, \swordfish uses a non-linear model for the synaptic device states (\Wcircled{3}) to map the weight matrices of digital weights into ideal corresponding conductance matrices (\Wcircled{4}). After that, \swordfish applies to these metrics the synaptic variations for the crossbar (\Wcircled{5}) that are determined from an analytical model based on the estimated behavior of memristor devices within a crossbar array. The output consists of the same number of matrices, but now with adjusted weights (\Wcircled{6}). \swordfish finally applies to those matrices the profile of all known circuit-level non-idealities (\Wcircled{7}) by adding representative metrics for these non-idealities. The output consists of matrices accounting for all variations and non-idealities (\Wcircled{8}).

\subsection{\accuracymitigator} \label{subsec:accuracy_mitigator}

Since accuracy is a critical metric in basecalling, \swordfish applies several mitigation techniques to deal with the non-idealities and their induced errors on the \vmm and/or basecalling. More specifically, \swordfish supports four different accuracy enhancement techniques: (1) analytical \vatlong (\vat) (offline), (2) \kdlong (\kd) training, (3) read–verify–write (R-V-W) training, and (4) \rsalong (\rsa) retraining (online).

\subsubsection{Analytical Variation-Aware Offline Training.}
\label{subsec:analytical_variation_aware_offline_retraining-mitigation_techniques_for_accuracy}

\swordfish supports \vatlong (\vat)~\cite{long2019design-variationAwareTrainingVAT1, liu2015vortex-variationAwareTrainingVAT2, chen2017accelerator-variationAwareTrainingVAT3, klachko2019improving-variationAwareTrainingVAT4}  during the training of a target \dnn as the simplest method to enhance the accuracy loss due to (1) quantization and (2) possible resistance variations per weight, which can be analytically or experimentally measured. Existing works randomly inject faults into the weights of the \dnn~\cite{markusfritscher2021simulating}, or model the potential errors at the end of each layer~\cite{long2019design-variationAwareTrainingVAT1, markusfritscher2021simulating}. Similarly, \swordfish utilizes the crossbar characterization for the errors per \vmm (i.e., the error library in the first approach in \vmmmodelgenerator) or an analytical crossbar model for the errors per \vmm (i.e., as in the second approach in \vmmmodelgenerator). \swordfish injects the modeled errors in the training and considers the rest of the devices unaltered. \swordfish repeats this process for each \vmm and every layer and then retrains the basecaller network. This way, \swordfish ensures that its retraining yields a better estimate for the errors arising from non-idealities in the crossbar.

\subsubsection{\kdlongcapital-based Variation-Aware Training.}
\label{subsec:kd_variation_aware_offline_retraining-mitigation_techniques_for_accuracy}

In addition to offline \vat based on injecting random errors or potential errors per layer discussed in \sect{\ref{subsec:analytical_variation_aware_offline_retraining-mitigation_techniques_for_accuracy}}, \swordfish is capable of supporting the \kdlong (\kd) approach as a \vat as well, i.e., \swordfish exploits knowledge/weights that exist in an ideal (typically a FP32-based) basecaller baseline to guide the training of \accswordfish, our memristor-based \cim design for \bonito. In \kd, two models exist: (1) the teacher (an ideal implementation using high precision data format, e.g., FP32-bit format) and (2) the student (\accswordfish quantized to 16-bit-width fixed-point presentation for both weights and activations). The goal is to mimic the teacher's output in the student by minimizing a loss function where the target is the result of applying the softmax on the quantile function associated with the standard logistic distribution (i.e., logit) of the teacher's training~\cite{hinton2015distilling-KnowledgeDistilationKD}. We refer the reader to previous works on \kd~\cite{hinton2015distilling-KnowledgeDistilationKD, charan2020accurate-AccurateInferenceInaccurateMemory} for further detail on how a loss function can be implemented in such a system to minimize the difference of \accswordfish's output and the teacher model's softmax output.

\subsubsection{\rvwlongcapital (\rvw) Training.}
\label{subsec:rvw_online_retraining-mitigation_techniques_for_accuracy}

\rvwlongcapital (\rvw) is a conventional error mitigation technique for non-ideal memristor-based memories that provides cell-by-cell error compensation. \rvw is used in \oldlong (\old)~\cite{liu2014reduction-RVWretraining2-openLoopOffDeviceOLD-AccurateInferenceInaccurateMemory17} where \rvw programming and sensing loop help the actual resistance of the device to converge to the expected target resistance. This method involves many read-and-write operations and feedback control for memristors, making \rvw a slow technique to mitigate accuracy loss. Note that to improve the accuracy in \rvw, we need to increase the fraction of the retrained weights (memristor devices in our case), increasing the cost of the mitigation technique. 

\subsubsection{\rsalongcapital Online Retraining.}
\label{subsec:rsa_online_retraining-mitigation_techniques_for_accuracy}

\swordfish uses \rsalong (\rsa)~\cite{charan2020accurate-AccurateInferenceInaccurateMemory} to map the learned \dnn model to \accswordfish. \rsa is used to mitigate the performance overhead of R-V-W technique~\cite{hu2013bsb-RVWretraining1, liu2014reduction-RVWretraining2-openLoopOffDeviceOLD-AccurateInferenceInaccurateMemory17}. \rsa by itself prevents only some of the non-idealities from being materialized as inaccuracies and can be an offline mechanism. However, \accswordfish combines it with an online training mechanism.

For its online retraining using \rsa, \swordfish places a small on-chip \sram-based memory next to memristor-based crossbars and distributes the learned \dnn model (i.e., weights) between this \sram and memristor-based crossbars. The key idea \swordfish uses is to map the weights that otherwise would map to error-prone memristor devices to reliable \sram cells. If one has access to the exact profile of the underlying memristor-based memory crossbars, one can exploit the knowledge on which memristors and columns are more error-prone and use this knowledge to decide which weight to map into the crossbar and which one to the \sram. In our evaluations of \swordfish, we use this knowledge whenever we use the chip measurements already used in the first approach of the \vmmmodelgenerator. However,  \swordfish can also randomly choose memristor devices in the crossbar and map (i.e., hardwire) them to the \sram. Random choice is the next best option without knowledge about the exact error pattern of a memristor-based crossbar. We used this method whenever we used the second approach (i.e., analytical model) in the \vmmmodelgenerator (\sect{\ref{subsec:vmm_model_generator}}).

\fig{\ref{fig:KDbased_weight_mapping_and_rsalong-variation_aware_offline_retraining-mitigation_techniques_for_accuracy}} presents how \accswordfish adopts \rsa with an online retraining mechanism (e.g., \kd) in a three-step approach:

\begin{figure}[htbp]
\centering
\setcounter{figure}{5}
    \includegraphics[width=1\linewidth]{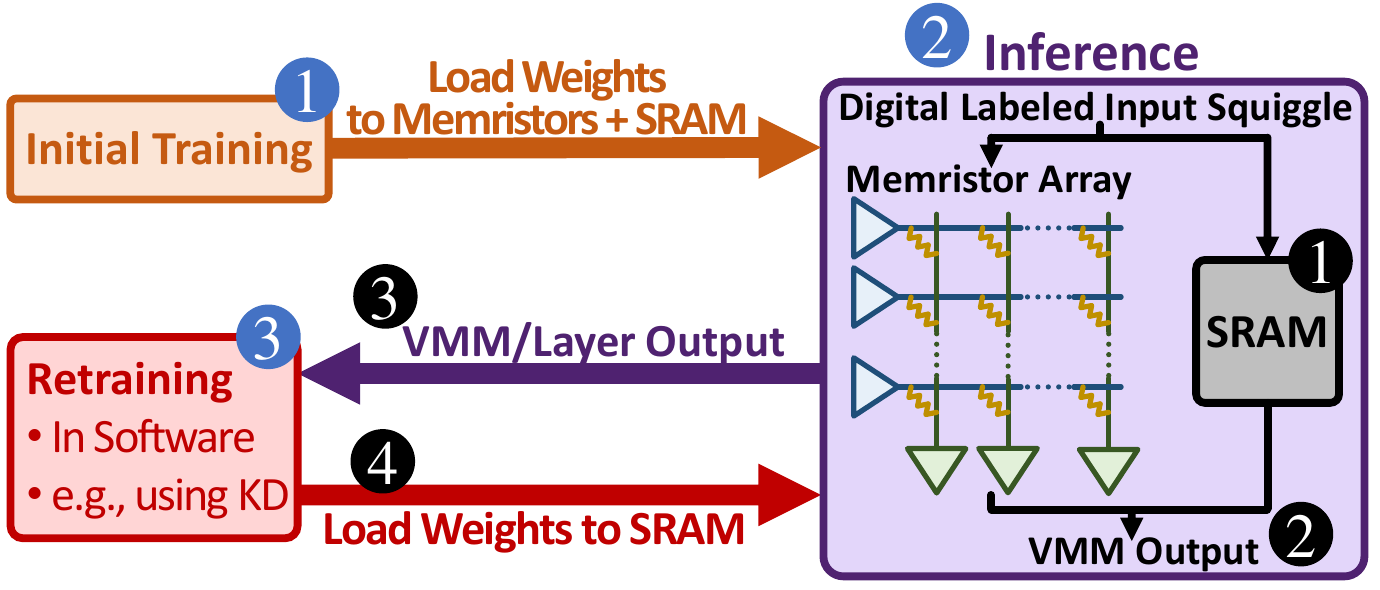}
    \caption{\swordfish's online error mitigation via \rsa.} 
    
    \label{fig:KDbased_weight_mapping_and_rsalong-variation_aware_offline_retraining-mitigation_techniques_for_accuracy}
\end{figure}

\begin{enumerate} [leftmargin=*]

    \item In the first step (\Bcircled{1}), \accswordfish trains the original \bonito and loads the initial weights from the \bonito \dnn model into the assigned memristor crossbar and the \sram (\circled{1}). \accswordfish considers this model as the initial model for the student in \kd.

    \item In the second step (\Bcircled{2}), \accswordfish performs a \vmm operation as usual. However, whenever one or more of the assigned weights to \sram (i.e., error-prone memristors or randomly chosen ones in \swordfish) is involved, \accswordfish reads the value from the \sram memory instead of the memristor device. \swordfish does this by passing the inputs of corresponding devices through the \sram value instead of the crossbar, zeroing the input for that particular memristor in the crossbar, and then summing up the values of both paths (\circled{2}).

    \item In the third step (\Bcircled{3}), \accswordfish returns the results of the \vmm operation of each crossbar (\circled{3}) to the retraining component (\kd in our example in  \fig{\ref{fig:KDbased_weight_mapping_and_rsalong-variation_aware_offline_retraining-mitigation_techniques_for_accuracy}}) and performs online training on only the weights that are mapped to the \sram memory to improve the accuracy loss due to non-idealities. Note that \accswordfish considers the non-ideality models of crossbars, \adc{}s, and \dac{}s to the student model for every training batch and trains the student. This includes both the initial training in Step \Bcircled{1} and retraining in Step \Bcircled{3}.

    \item \accswordfish then loads the new weights to the \sram near the crossbars (\circled{4}) and repeats Steps~\Bcircled{2} and \Bcircled{3}. 

\end{enumerate}

\accswordfish uses \kd-based variation aware training for its Step~\circled{3} in \fig{\ref{fig:KDbased_weight_mapping_and_rsalong-variation_aware_offline_retraining-mitigation_techniques_for_accuracy}} online retraining. However, any other retraining method can also replace \kd in our example. Note that all the parameters are already quantized to 16-bit fixed-point precision to present the model in \accswordfish accurately. \swordfish leverages the weights from the converged teacher model to improve the convergence of the student model.

\rsa in \swordfish comes at the price of extra area overhead for the considered on-chip \sram memory, storage in the memory controller for mapping metadata, summation of the output from the crossbar with on-chip memory, and some additional control logic evaluated in \sect{\ref{sec:evaluations_notSmart_and_smartMitigations}}.

\subsection{\systemevaluator} \label{subsec:system_evaluator}

The \systemevaluator module puts the results of all previous modules of \swordfish together to evaluate the target \dnn.

As inputs, this module takes the execution time for each \vmm operation, the accuracy of each \vmm operation for the last layer of the \dnn (as it determines the final accuracy of the \dnn), the number of active crossbars in each step of \swordfish, and information in peripheral circuitry. 

The \systemevaluator module has 3 outputs: 

\begin{enumerate} [leftmargin=*]

    \item \textbf{Accuracy:} The \systemevaluator module outputs an accuracy number for the evaluated \dnn. In \accswordfish, this number shows the accuracy of the basecaller, commonly known as \emph{read accuracy}, which is the fraction of the total number of exactly matching bases of a read to a reference to the length of their alignment (including insertions and deletions).

    \item \textbf{Basecalling throughput:}  The \systemevaluator module outputs a number for inference throughput of the target \dnn. In \accswordfish, this number is the basecalling throughput, defined as kilo-basepairs generated by the basecaller per second ($\frac{Kbp}{s}$). The higher the basecalling throughput, the better. This is the most important metric to evaluate a basecalling accelerator's performance. Our throughput evaluations in \accswordfish include the time required for read and write time for the inputs and outputs, respectively.\footnote{We use this command line in Linux: /usr/bin/time -v.}

    \item \textbf{Area overhead.} The \systemevaluator module of \swordfish also reports area overhead based on the underlying architecture to account for the overheads of a dedicated accelerator, e.g., \accswordfish. 
    
\end{enumerate}

\subsection{\swordfish Evaluation Challenges} \label{sec:swordfish_in_realWorld}

Comprehensive, fair, and practical evaluation of \swordfish is challenging for two main reasons. First, most of the \sota basecallers are either not open-source~\cite{lou2020helix, huang2020sacall, wick2019performance-Guppy} or support only specific reads~\cite{dorado2022}. Second, current simulators and frameworks mimicking memristor-based \cim designs are either not open-source, do not consider the underlying non-idealities of the devices, or only support a very limited number of non-idealities, emerging technologies, or neural networks~\cite{lin2018dl-dlrsimMarkus8rdRefofSimulLASCAS, jain2020rxnn-NonIdealities-ADCDACVariations}.

To evaluate \swordfish despite these challenges, we take two representative examples. Specifically, for the first challenge, we primarily compare our method with \bonito~\cite{bonito2020}, an open-sourced, universally applicable tool currently under active development and maintenance by \ont (\sect{\ref{subsec:genome_sequencing_pipeline-Background}}). \bonito stands out for its exceptional accuracy and performance over its predecessors like \guppy~\cite{wick2019performance-Guppy} and does not face the limited support for reads (e.g., \dorado~\cite{dorado2022}) or lack of open-source implementation and training code (e.g., \helix~\cite{lou2020helix}, \halcyon~\cite{konishi2021halcyon}, \guppy~\cite{wick2019performance-Guppy}, and \sacall~\cite{huang2020sacall}). For the second challenge, we consider \puma architecture as the baseline architecture for the two reasons mentioned in \sect{\ref{subsec:programmable_accelerator_architectures_memristorbased_PUMA_ISAAC-Background}}.

\section{Evaluation Methodology} 
\label{sec:experimental_setup_and_methodology}

\subsection{Implementations and Models} \label{subsec:implementation-experimental_setup_and_methodology}

For the performance and area studies, we significantly extended the \puma simulator and \puma compiler to account for (1) \bonito's \dnn architecture, (2) updated configurations in Core Architecture of \puma~\cite{ankit2019puma} based on our memory models and the TSMC \SI{40}{\nano\meter}~\cite{kim2016multistate-hfo2tiox} technology node used for peripheries, and (3) performance and area overheads introduced by non-idealities of memristors and their mitigation techniques. Note that we use Synopsys Design Compiler~\cite{synopsys} and synthesize the additional components of our design in the target technology to obtain their execution time, power, and area. We apply the prominent technology scaling rules~\cite{sarangi2021deepscaletool} to the configuration numbers of the \puma architecture to ensure all of our design components are based on the same technology node.

For accuracy analysis (in both training and inference phases), we also extensively modified \bonito's open-source implementation~\cite{bonito2020} to consider the device characteristics and limitations of the architecture. Unfortunately, \puma does not allow us for such analysis as it considers the effects of only quantization and write variations on accuracy.

We utilize prototyped cross-array memristors as our memory arrays and capture the variations in their spatiotemporal conductivity, execution time, and area overhead of necessary operations. We project our characterization results of real memories to our \dnn evaluations. We also build a statistical model from our measurements to capture the full picture of a larger memory model for large-scale variations, timing, and area parameters. This model contains four types of variations: (1) input DACs, (2) synaptic variations, (3) wire resistance, and  (4) output ADCs. The memory prototypes and models used for evaluations and simulations are based on the results of the EU project MNEMOSENE~\cite{MNEMOSENE-TUDelftEU}, concluded in 2020, generously provided by the involved parties. The results have been tested heavily during the project and by various metrics found in the related literature. \tab{\ref{tab:memristorparameters-experimental_setup_and_methodology}} shows the main parameters of our memristor-based crossbars.

\begin{table}[htbp]
\vspace{0.5em}
\small
\scriptsize
\centering
\begin{tabular}{l|l}

\hline
\textbf{Technology and device} & \reram $HfO_2 / TiO_x$~\cite{kim2016multistate-hfo2tiox}\\ \hline
\textbf{Cell configuration} & 1T1R (NMOS T: \SI{460}{\nano\meter}/\SI{40}{\nano\meter}\\ \hline
\textbf{HRS/LRS} & \SI{1}{\mega\ohm}/\SI{10}{\kilo\ohm}\\ \hline
\textbf{n$_{min}$/n$_{max}$} & 0.03, 30\\ \hline
\textbf{Array Sizes} & 64$\times$64 and 256$\times$256\\ \hline
\textbf{SA V$_{min}$} & \SI{40}{\milli\volt}\\ \hline
\end{tabular}
\caption{Our array and device configurations.}
\label{tab:memristorparameters-experimental_setup_and_methodology}
\end{table}

Our study specifically evaluates \swordfish on \reram memristors for three reasons. First, the availability of actual chip measurements is essential for our non-ideality-centered study. Second, lower energy costs for writing/programming than alternatives like \pcm. Third, \reram's established status within the memristor family provides reliable baselines and intuitions for device-level features, enhancing the credibility of our proposal.

\subsection{Simulation Infrastructure} \label{subsec:simulation_infrastructure-experimental_setup_and_methodology}

We ran our baseline \bonito basecaller and software implementation of \swordfish on a 128-core server with AMD EPYC 7742 \cpu{}s~\cite{AMDEPYC7742CPU}, 500GB of DDR4 DRAM, and 8 NVIDIA V100~\cite{NVIDIAV100} cards. We train and evaluate \swordfish accuracy and software results on our NVIDIA cards (with 32-bit floating-point precision). We use the nvprof profiler~\cite{yang2020hierarchical-nvprofinBaro} for the profiling experiments on \gpu.

\subsection{Evaluation Metrics} \label{subsec:metrics-experimental_setup_and_methodology}

We use metrics output by the \systemevaluator module for our comparisons. \sect{\ref{subsec:system_evaluator}} clarifies these metrics.

\subsection{Datasets and Workloads} \label{subsec:datasets_and_workloads-experimental_setup_and_methodology}

\tab{\ref{tab:datasets_and_workloads-experimental_setup_and_methodology}} provides datasets from a MinION R9.4.1 flowcell~\cite{ReadSetBasecalling, ReferenceGenomeSetBasecalling} we use in our evaluations.

\begin{table}[htbp]
\vspace{0.5em}
\small
\scriptsize
\centering
\renewcommand{\tabcolsep}{1pt}
\begin{tabular}{|c c||c|c|}

\hline
 & Dataset (Organism)                                                                     & \# Reads & Reference Genome Size (bp) \\\hline \hline
D1             & \begin{tabular}[c]{@{}c@{}}Acinetobacter pittii\\ 16-377-0801\end{tabular}   & 4,467    & 3,814,719                                                             \\\hline
D2             & \begin{tabular}[c]{@{}c@{}}Haemophilus haemolyticus\\ M1C132\_1\end{tabular} & 8,669    & 2,042,591                                                             \\\hline
D3             & \begin{tabular}[c]{@{}c@{}}Klebsiella pneumoniae\\ NUH29\end{tabular}        & 11,047   & 5,134,281                                                             \\\hline
D4             & \begin{tabular}[c]{@{}c@{}}Klebsiella pneumoniae\\ INF042\end{tabular}       & 11,278   & 5,337,491                     \\\hline
\end{tabular}
\caption{Read and Reference Datasets for our Basecalling Evaluation.}
\label{tab:datasets_and_workloads-experimental_setup_and_methodology}
\end{table}

\section{\swordfish Evaluation} 
\label{sec:evaluations_notSmart_and_smartMitigations}

We first use \swordfish to investigate the impact of constraints and non-idealities of a \puma-based architecture (\sect{\ref{subsec:memristors_noises-Background}}) on the accuracy of the \bonito basecaller~\cite{bonito2020}. We call this design the \nomitigationhwaccswordfish, as it achieves the highest performance for our memristor-based hardware accelerator without any accuracy enhancement technique. We then explore the effect of the accuracy enhancement mechanisms in \swordfish applied to deal with the inaccuracies of the memristor-based accelerator as it affects the \bonito basecaller's accuracy. The results of this design are presented under \realhwaccswordfish.

\subsection{Effect of Quantization on Accuracy without Accuracy Enhancement}
\label{subsec:accuracy_quantization_effects-evaluations_notSmart}

Since both the weights and activations in the original \dnn are in FP32 format, \swordfish can opt for quantizing one or both of them. The degree of the quantization can differ depending on how much each parameter impacts the overall accuracy. \swordfish considers seven different configurations: the default configuration (DFP 32-32), where weights and activations use the FP32\footnote{FP stands for floating point.} format, and 6 FPP X-Y\footnote{FPP stands for fixed point precision.} formats, where X and Y denote the fixed-point precision of weights and activations, respectively. \swordfish currently only supports power-of-two precision levels for its quantized configurations. \tab{\ref{tab:accuracy_after_quantizations_Bonito_LSTM_Architecture-evaluations_notSmart}} presents the accuracy of different configurations.

\begin{table}[htbp]
\vspace{0.5em}
\small
\scriptsize
\centering
\renewcommand{\tabcolsep}{1pt}
\begin{tabular}{|c||c|c|c|c|c|c|c|}

\hline
\textbf{} & \textbf{DFP 32-32} & \textbf{FPP 16-16} & \textbf{FPP 8-8} & \textbf{FPP 8-4} & \textbf{FPP 4-8} & \textbf{FPP 4-4} & \textbf{FPP 4-2} \\ \hline \hline
\textbf{D1}                                                                        & 97.32\%              & 97.32\%              & 97.12\%            & 97.12\%            & 95.42\%            & 95.62\%            & 93.62\%            \\ \hline
\textbf{D2}                                                                        & 97.32\%              & 97.32\%              & 96.72\%            & 96.72\%            & 94.92\%            & 95.42\%            & 92.42\%            \\ \hline
\textbf{D3}                                                                        & 97.32\%              & 97.32\%              & 96.02\%            & 95.82\%            & 93.62\%            & 95.12\%            & 93.72\%            \\ \hline
\textbf{D4}                                                                        & 97.32\%              & 97.32\%              & 96.42\%            & 96.42\%            & 94.22\%            & 95.32\%            & 93.62\%            \\ \hline
\end{tabular}
\caption{Accuracy evaluation after quantization.}
\label{tab:accuracy_after_quantizations_Bonito_LSTM_Architecture-evaluations_notSmart}
\end{table}

We make two major observations. First, \bonito's architecture can tolerate some quantization level without accuracy loss. More specifically, across all evaluated datasets, quantization down to 16 bits does not affect the accuracy at all, and quantization down to 8 bits reduces the accuracy by less than \maxAccuracyDropQuantizationNoSmart even in extreme cases. We conclude that \nomitigationhwaccswordfish can still reduce the precision of its network from a 32-bit FP format to 16-bit-width fixed point precision without accuracy loss. This way, \nomitigationhwaccswordfish can (1) accelerate the network on a platform limited to fixed point format representation and (2) improve the energy efficiency of the network via lower data precision. This observation is on par with similar studies~\cite{shafiee2016isaac, singh2022framework-RUBICONQABASSkipClipRUBICALL-gaganbasecaller, jain2019cxdnn} exploiting quantization as a technique to improve the performance and energy efficiency of a \dnn with a negligible accuracy loss.

Second, tolerance to quantization varies depending on the input dataset. This makes the effect of quantization on accuracy workload-dependent. However, the accuracy drop for different quantization configurations follows more-or-less a similar trend irrespective of the dataset, i.e., they all follow a decreasing trend with reduced data representation. We conclude that \swordfish's understudy network (\bonito) tolerates some quantization but will offer very low accuracy for extreme quantization (i.e., lower than 4-bit precision) irrespective of the dataset. We note that an accuracy drop of $\sim$5\% and higher is considered unacceptable for a future basecaller, as accuracy is the most critical metric in \sota basecallers. This observation is consistent with prior works on smaller~\cite{jain2020rxnn-NonIdealities-ADCDACVariations} or different types of networks~\cite{singh2022framework-RUBICONQABASSkipClipRUBICALL-gaganbasecaller}.

We conclude that quantization is a viable solution to tackle data representation constraints in hardware accelerators and, therefore, can be used in a framework such as \swordfish. However, accuracy loss due to quantization (applied with the expectance of accuracy loss due to variations and non-idealities) leads us to consider only down to 16 (or possibly 8) bits of precision for both weights and activations before a significant accuracy drop occurs. Therefore, the following studies consider only a 16-bit integer as the quantization level.

\subsection{Effect of Non-idealities on Accuracy without Accuracy Enhancement}
\label{subsec:accuracy_nonidealities_effects-evaluations_notSmart}

We examine the effect of four non-idealities on basecalling accuracy. The results presented in this section belong to the second approach of modeling non-idealities in the \vmmmodelgenerator module, i.e., using analytical modeling (see \sect{\ref{subsec:vmm_model_generator}}).

\subsubsection{Effect of Write Variation on Accuracy.}
\label{subsec:write_variation_effect_on_Accuracy-accuracy_nonidealities_effects-evaluations_notSmart}

Write variation can single-handedly impact the accuracy results of a \vmm operation~\cite{jain2019cxdnn, charan2020accurate-AccurateInferenceInaccurateMemory}. Therefore, we analyze it separately.

\fig{\ref{fig:accuracy_after_writevariation_Bonito_LSTM_Architecture-evaluations_notSmart}} presents the effects of write variations on accuracy. The x-axis sweeps the write variation rate. The error bars account for the accuracy variations on different write variation rates over 1000 runs of the model. Since the models for write variation are circuit-dependent and have varying probabilities of affecting the stored/programmed data, this methodology provides us with a better insight into the effect of this non-ideality on accuracy.

\begin{figure}[htbp]
\centering
\setcounter{figure}{6}
    \includegraphics[width=1\linewidth]{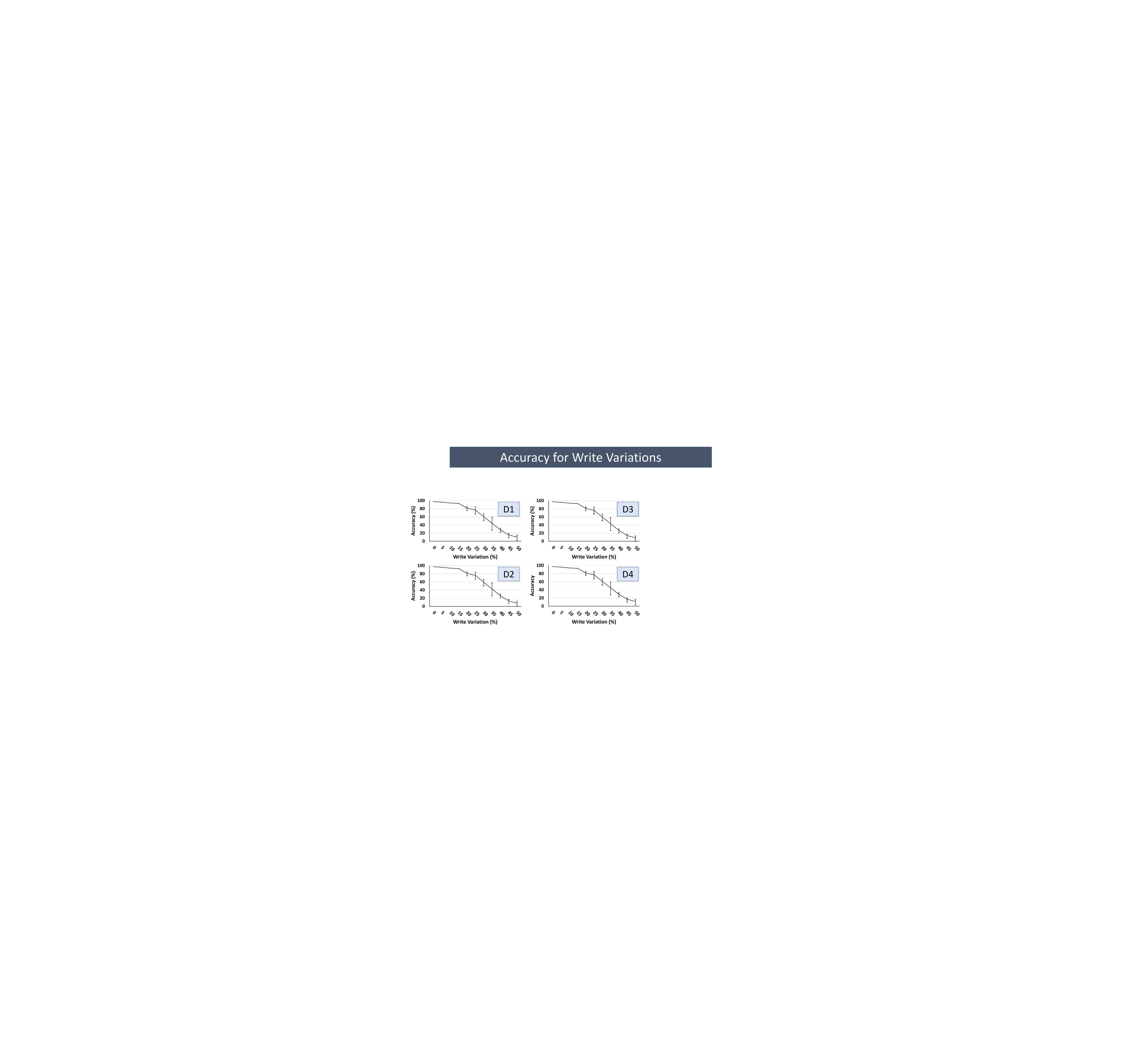}
    \caption{Accuracy after taking into account write variation.}

    \label{fig:accuracy_after_writevariation_Bonito_LSTM_Architecture-evaluations_notSmart}
\end{figure}

\begin{figure*}[!htbp]
    \centering
    \setcounter{figure}{7}
    \begin{minipage}{1\textwidth}
        \centering
        \includegraphics[width=1\textwidth]{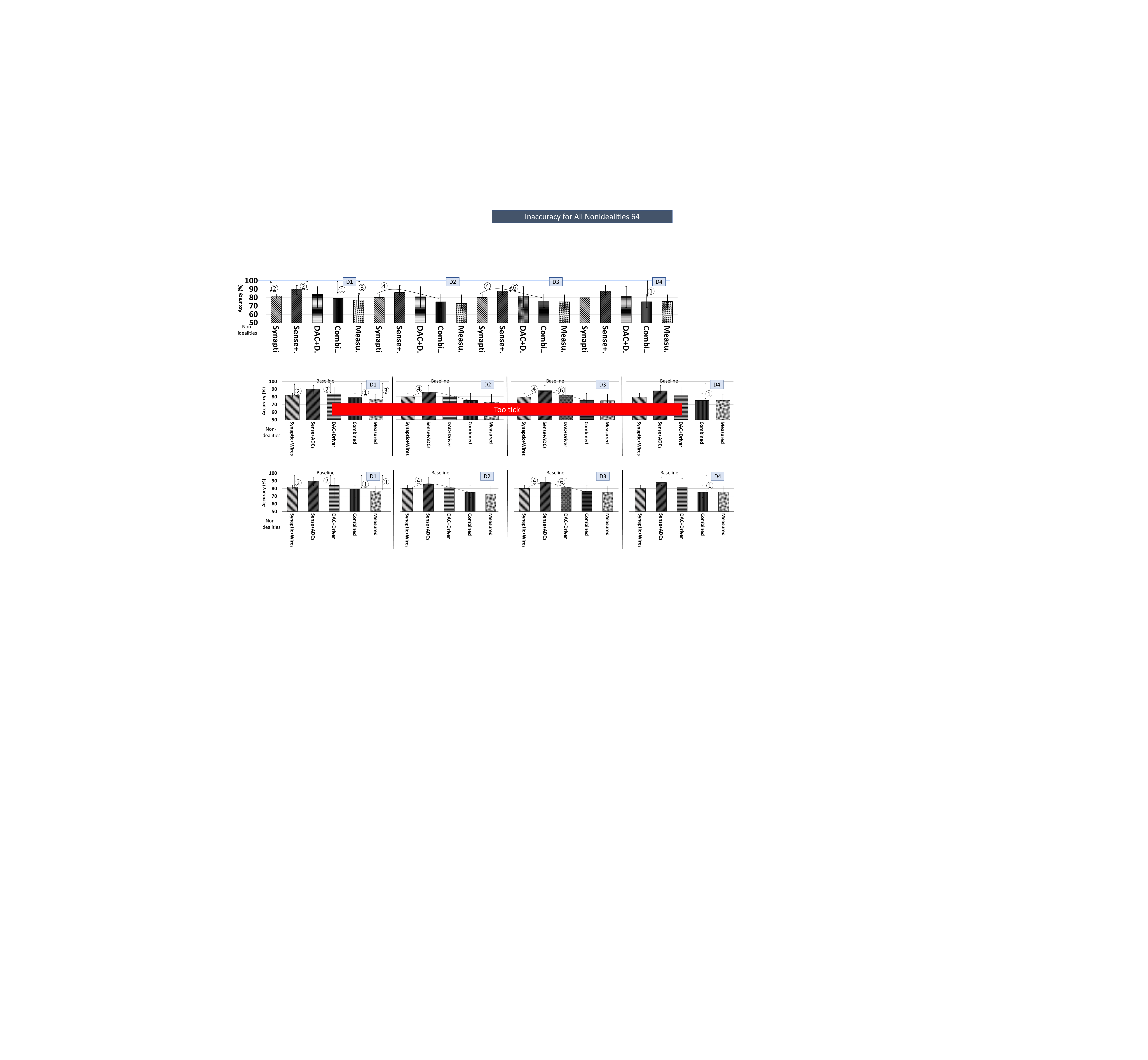} 
        \caption{Accuracy after taking into account non-idealities on 64$\times$64 crossbars for the 4 datasets.}
        \label{fig:accuracy_after_allNonIdealities64_Bonito_LSTM_Architecture-evaluations_notSmart}
    \end{minipage}\hfill
    \begin{minipage}{1\textwidth}
        \centering
        \setcounter{figure}{8}
        \includegraphics[width=1\textwidth]{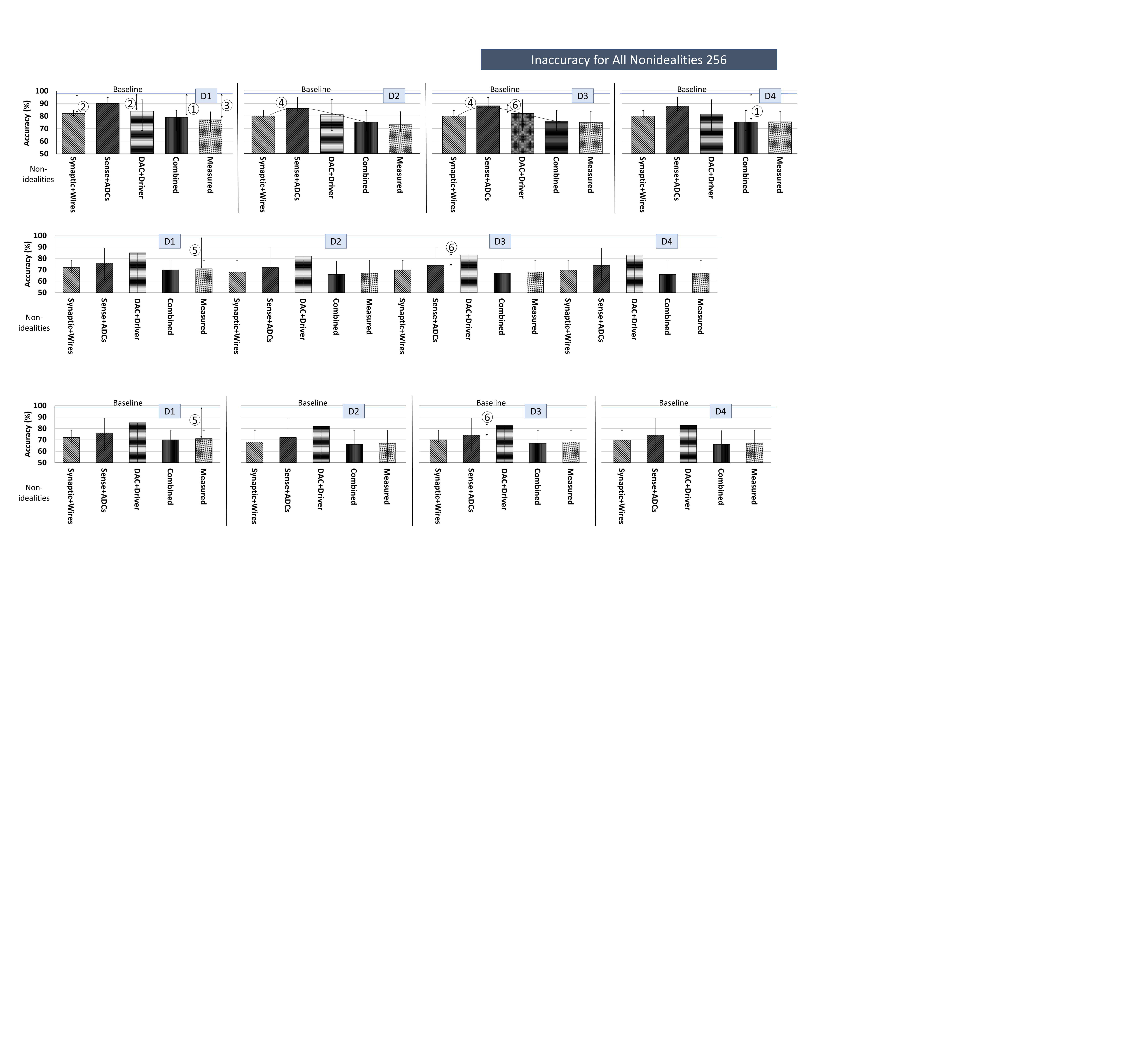} 
        \caption{Accuracy after taking into account non-idealities on 256$\times$256 crossbars for the 4 datasets.} 
        
        \label{fig:accuracy_after_allNonIdealities256_Bonito_LSTM_Architecture-evaluations_notSmart}
    \end{minipage}\hfill
    \label{fig:accuracy_after_allNonIdealities_Bonito_LSTM_Architecture-evaluations_notSmart}
\end{figure*}

We make two main observations. First, slight write variation can lead to a significant drop in the accuracy of end-to-end basecalling. To a great extent, this is on par with previous works' observation of the write variation impact on \vmm accuracy~\cite{jain2019cxdnn, charan2020accurate-AccurateInferenceInaccurateMemory}. For example, the accuracy drops vary from \accuracyWriteVariationTenACINETOBACTER to \accuracyWriteVariationFiftyACINETOBACTER for D1 and from \accuracyWriteVariationTenKLEBSIELLA to \accuracyWriteVariationFiftyKLEBSIELLA for D4.

Second, the exact accuracy loss depends on the input dataset, i.e., the accuracy is workload-dependent and varies for the same write variation among different subfigures in \fig{\ref{fig:accuracy_after_writevariation_Bonito_LSTM_Architecture-evaluations_notSmart}}. For example, for the same write variation rate of \writevariationTwentyFive, the accuracy on our two datasets (i.e., D2 and D4) can vary by \accuracyDiffAvgWritevariationTwentyFiveHAEMOvsKLEBS.

We conclude that write variation in \nomitigationhwaccswordfish can debilitate the basecalling process significantly. In other words, write variation can eliminate all the potential performance and energy efficiency benefits of such a memristor-based design if not mitigated correctly. Therefore, unlike the quantization constraint, we should closely control the write variations in any future design for an acceptable basecaller. Fortunately, some previous works~\cite{pedretti2021conductance, markusfritscher2022mitigating, charan2020accurate-AccurateInferenceInaccurateMemory} propose mitigation techniques that, when combined, can provide us with reasonable (e.g., amount of $\le$ \maxMitigateWriteVariationAfterAllMitigations) write variation. From now on, we consider only up to 10\% write variation (as defined in \sect{\ref{subsec:memristors_noises-Background}}) in our evaluations.

\subsubsection{Effect of Combined Non-idealities on Accuracy.}
\label{subsec:combined_nonidealities-accuracy_nonidealities_effects-evaluations_notSmart}

\fig{\ref{fig:accuracy_after_allNonIdealities64_Bonito_LSTM_Architecture-evaluations_notSmart}} and \fig{\ref{fig:accuracy_after_allNonIdealities256_Bonito_LSTM_Architecture-evaluations_notSmart}} show the accuracy after considering all other sources of non-idealities (see \sect{\ref{subsec:memristors_noises-Background}}) for our four datasets on two different crossbar sizes of 64$\times$64 and 256$\times$256, respectively. The error bars show the distribution when considering \maxMitigateWriteVariationAfterAllMitigations write variation over 1000 runs. For each dataset, \fig{\ref{fig:accuracy_after_allNonIdealities64_Bonito_LSTM_Architecture-evaluations_notSmart}} and \fig{\ref{fig:accuracy_after_allNonIdealities256_Bonito_LSTM_Architecture-evaluations_notSmart}} present the accuracy results for five configurations presented as individual bars in the figures. The first three bars from the left present the results for individual non-idealities, i.e., synaptic+wire resistances (\emph{Synaptic+Wires}), sensing+\adc{} circuitry (\emph{Sense+\adc}), and \dac{}+driver circuitry (\emph{\dac{}+Driver}), respectively, that \swordfish accounts for in its second approach of modeling non-idealities in the \vmmmodelgenerator module, i.e., using analytical modeling (\sect{\ref{subsec:vmm_model_generator}}). The fourth bar, \emph{Combined}, accounts for all the non-idealities from the same analytical model simultaneously. The fifth and last bar, \emph{Measured}, considers all the non-idealities from the library of real chip measurements in the first approach of modeling non-idealities in the \vmmmodelgenerator (see \sect{\ref{subsec:vmm_model_generator}}).\footnote{We leave the exploration of every possible combination of individual non-idealities to future work.} We make six main observations.

\begin{figure*}[b]
\centering
\setcounter{figure}{10}
    \includegraphics[width=1\linewidth]{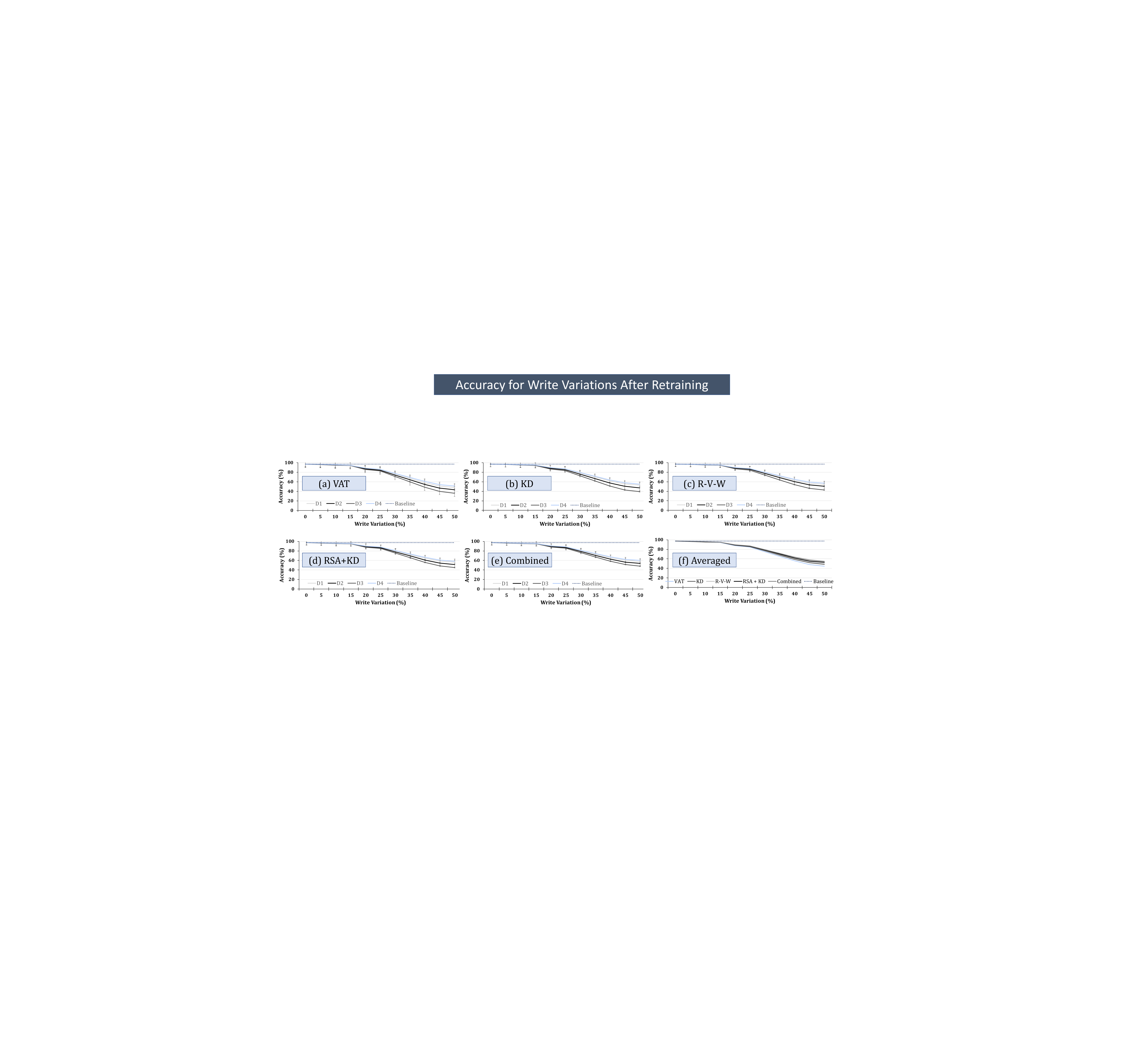}
    \caption{Accuracy after combining enhancement techniques over different write variations.}
\label{fig:accuracy_after_writevariation_Bonito_LSTM_Architecture_With_Mitigation-evaluations_smart_mitigations}
\end{figure*}

\begin{enumerate} [leftmargin=*]
    \item A combination of non-idealities (i.e., each of the bars labeled with "Combined" or "Measured" or the 4th and the 5th bar per dataset in \fig{\ref{fig:accuracy_after_allNonIdealities64_Bonito_LSTM_Architecture-evaluations_notSmart}} and \fig{\ref{fig:accuracy_after_allNonIdealities256_Bonito_LSTM_Architecture-evaluations_notSmart}}) leads to a significant accuracy loss irrespective of the dataset or crossbar size. For example, observe the accuracy loss when considering all the non-idealities in an analytical way (bars labeled as "Combined"). The accuracy loss varies from \accuracyDropNoMitigationAccswordfishVsBonitoMin to \accuracyDropNoMitigationAccswordfishVsBonitoMax (\Wcircled{1} in \fig{\ref{fig:accuracy_after_allNonIdealities64_Bonito_LSTM_Architecture-evaluations_notSmart}}) across different datasets (i.e., D1 to D4). The same trend can be observed in \fig{\ref{fig:accuracy_after_allNonIdealities256_Bonito_LSTM_Architecture-evaluations_notSmart}}.

    \item The impact of individual non-idealities (i.e., \emph{Synaptic+Wires}, \emph{Sense+\adc}, or \emph{\dac{}+Driver}) on the accuracy (loss) is different. For example, observe the accuracy loss of \emph{\dac{}+Driver} versus \emph{Synaptic+Wires} in D1 (\Wcircled{2} in \fig{\ref{fig:accuracy_after_allNonIdealities64_Bonito_LSTM_Architecture-evaluations_notSmart}}). For the same dataset, the accuracy loss varies from \accuracyDropNoMitigationAccswordfishVsBonitoDACsAndDriversACINETOBACTER for \emph{\dac{}+Driver} to \accuracyDropNoMitigationAccswordfishVsBonitoWiresAndsynapticACINETOBACTER for \emph{Synaptic+Wires}. A similar difference also exists in crossbars of size 256$\times$256 in \fig{\ref{fig:accuracy_after_allNonIdealities256_Bonito_LSTM_Architecture-evaluations_notSmart}}.

    \item The accuracy loss for combined non-idealities is non-additive. For example, in D1, the total accuracy loss of \emph{Measured} is \accuracyDropNoMitigationAccswordfishVsBonitoAddsupACINETOBACTER (\Wcircled{3} in \fig{\ref{fig:accuracy_after_allNonIdealities64_Bonito_LSTM_Architecture-evaluations_notSmart}}) yet the simple addition of numerical accuracy loss of \emph{Synaptic+Wires}, \emph{Sense+\adc}, and \emph{\dac{}+Driver} totals \accuracyDropNoMitigationAccswordfishVsBonitoAllApproachTwoACINETOBACTER. We conclude that certain errors mask others.

    \item Accuracy loss values follow a similar trend irrespective of the dataset. See the trendlines \Wcircled{4} in \fig{\ref{fig:accuracy_after_allNonIdealities64_Bonito_LSTM_Architecture-evaluations_notSmart}} for D2 and D3. However, absolute accuracy loss values vary from one dataset to another. 

    \item The smaller the crossbar, the lower the accuracy loss. For example, for D1, we have lower accuracy loss (of \accuracyDropNoMitigationAccswordfishVsBonitoAllApproachOneACINETOBACTERSixtyfour versus \accuracyDropNoMitigationAccswordfishVsBonitoAllApproachOneACINETOBACTERTwohundredfiftysix) when using a 64$\times$64 crossbar compared to a  256$\times$256 crossbar (\Wcircled{3} in \fig{\ref{fig:accuracy_after_allNonIdealities64_Bonito_LSTM_Architecture-evaluations_notSmart}} vs. \Wcircled{5} in \fig{\ref{fig:accuracy_after_allNonIdealities256_Bonito_LSTM_Architecture-evaluations_notSmart}} for the \emph{Measured} configuration). This is because a smaller crossbar has mostly smaller accumulative noise induced in wires of a smaller array.

    \item Different non-idealities affect the same dataset differently for different crossbar sizes. For example, the accuracy loss due to non-idealities in \emph{\dac{}+Driver} is more dominant than those in \emph{Sense+\adc} on a 64$\times$64 crossbar, while this is the opposite for a 256$\times$256 crossbar. See \Wcircled{6} in \fig{\ref{fig:accuracy_after_allNonIdealities64_Bonito_LSTM_Architecture-evaluations_notSmart}} and \fig{\ref{fig:accuracy_after_allNonIdealities256_Bonito_LSTM_Architecture-evaluations_notSmart}}.
\end{enumerate}

Even for small yet practical crossbars of size 64$\times$64, the accuracy loss observed in this section under both \emph{Combined} and \emph{Measured} configurations in \fig{\ref{fig:accuracy_after_allNonIdealities64_Bonito_LSTM_Architecture-evaluations_notSmart}} and \fig{\ref{fig:accuracy_after_allNonIdealities256_Bonito_LSTM_Architecture-evaluations_notSmart}} is still significant (e.g., from \accuracyDropNoMitigationAccswordfishVsBonitoCombinedSixtyfourApproachOneMin to \accuracyDropNoMitigationAccswordfishVsBonitoMeasuredSixtyfourApproachOneMin) and unacceptable for a basecalling step that affects many other steps of a genome sequencing pipeline. We conclude that non-idealities in the memristor-based \cim designs, especially when combined, can be detrimental to basecalling accuracy and must be accounted for and mitigated before considering such a design useful in any other aspect.

\subsection{Effect of Accuracy Enhancement on Quantized Basecallers}
\label{subsec:accuracy_mitigations_effects_quantization-evaluations_smart_mitigations}

\fig{\ref{fig:accuracy_after_quantizationAwareRetraining_Bonito_LSTM_Architecture-accuracy_mitigations_effects_quantization-evaluations_smart_mitigations}} shows the results of applying \swordfish's accuracy enhancement techniques to a quantized \bonito basecaller. The x-axis presents six configurations for quantization as defined in \sect{\ref{subsec:accuracy_quantization_effects-evaluations_notSmart}}. For each quantization configuration, we evaluate five accuracy enhancement techniques, namely \emph{\vat}, \emph{\kd}, \emph{\rvw}, \emph{\rsa{}+\kd} (see \sect{\ref{subsec:accuracy_mitigator}}), and a combination of all techniques labeled as \emph{All}. The y-axis shows the accuracy of each technique for the corresponding quantization configuration. The horizontal line marked as Baseline (DFP 32-32) is the baseline accuracy as defined in \sect{\ref{subsec:accuracy_quantization_effects-evaluations_notSmart}}.

\begin{figure}[htbp]
\centering
\setcounter{figure}{9}
    \includegraphics[width=1\linewidth]{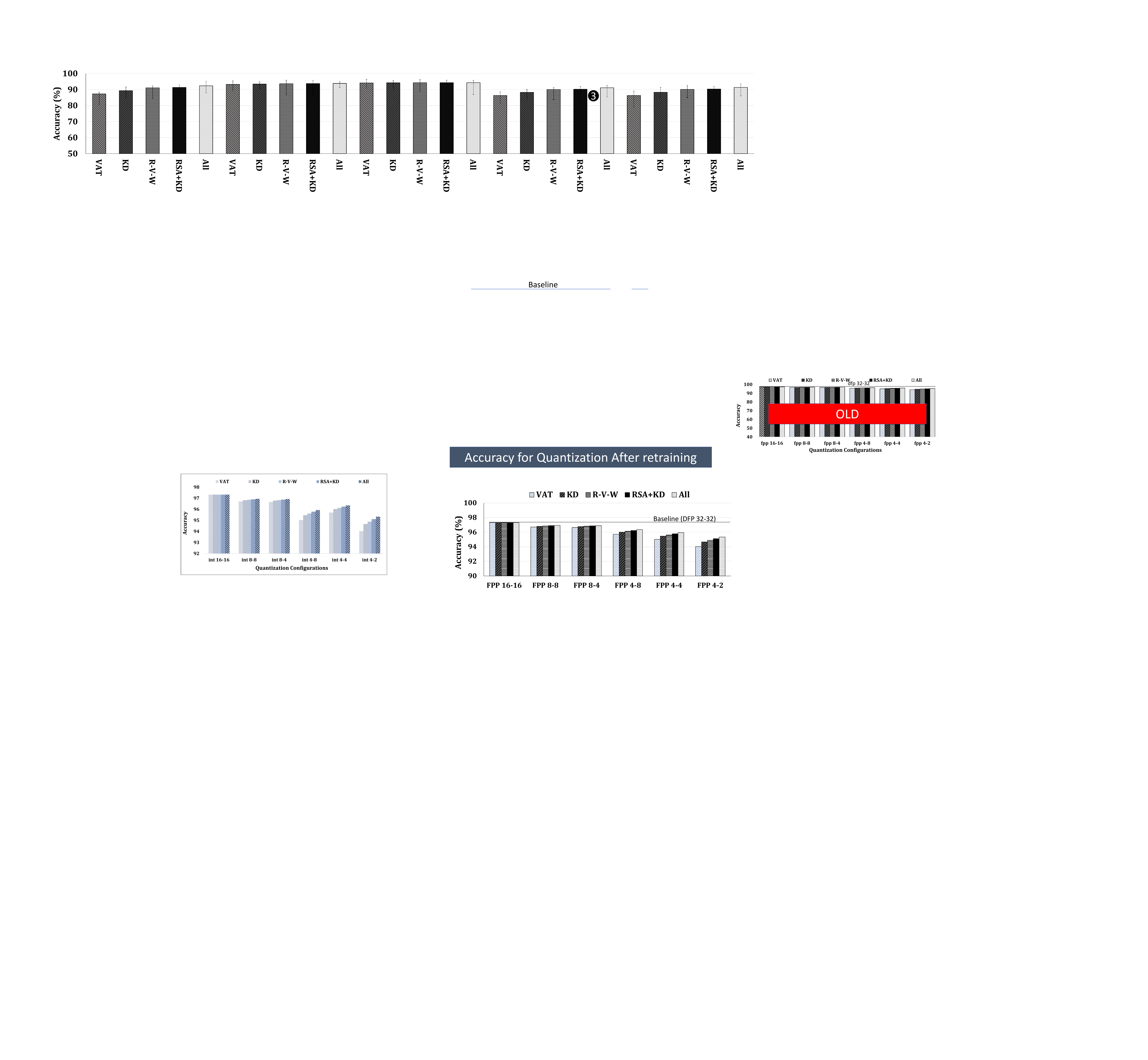}
    \caption{Accuracy enhancement after quantization.}
    \label{fig:accuracy_after_quantizationAwareRetraining_Bonito_LSTM_Architecture-accuracy_mitigations_effects_quantization-evaluations_smart_mitigations}
\end{figure}

We observe that retraining with quantization is an effective way to mitigate the accuracy loss induced by quantization. Our results show that with only \retraingIterationsQuantizationAverage extra retraining epochs, accuracy improves by \accuracyImprovementRetrainingQuantizationAwareAverage on average, for a basecaller quantized down to 8-bit. By applying all quantization-aware retraining methods that we discuss in \sect{\ref{subsec:accuracy_quantization_effects-evaluations_notSmart}}, \swordfish can retain the same accuracy as the \bonito basecaller with 32-bit floating point precision. This result is in agreement with the prior work on different types of neural networks~\cite{jain2020rxnn-NonIdealities-ADCDACVariations}. However, \swordfish is the first work to show this result for genomic basecalling. From now on, we use 16-bit precision quantization for all evaluations we show in the remainder of this paper. We conclude that the proposed mitigation mechanisms effectively mitigate the accuracy loss due to a reasonable amount of quantization, e.g., from 32-bit to 16-bit in the \bonito basecaller.

\subsection{Effect of Accuracy Enhancement on Non-idealities}
\label{subsec:accuracy_mitigations_effects_nonidealities-evaluations_smart_mitigations}

\begin{figure*}[!htbp]
    \centering
    \begin{minipage}{1\textwidth}
        \centering
        \setcounter{figure}{11}
        \includegraphics[width=1\textwidth]{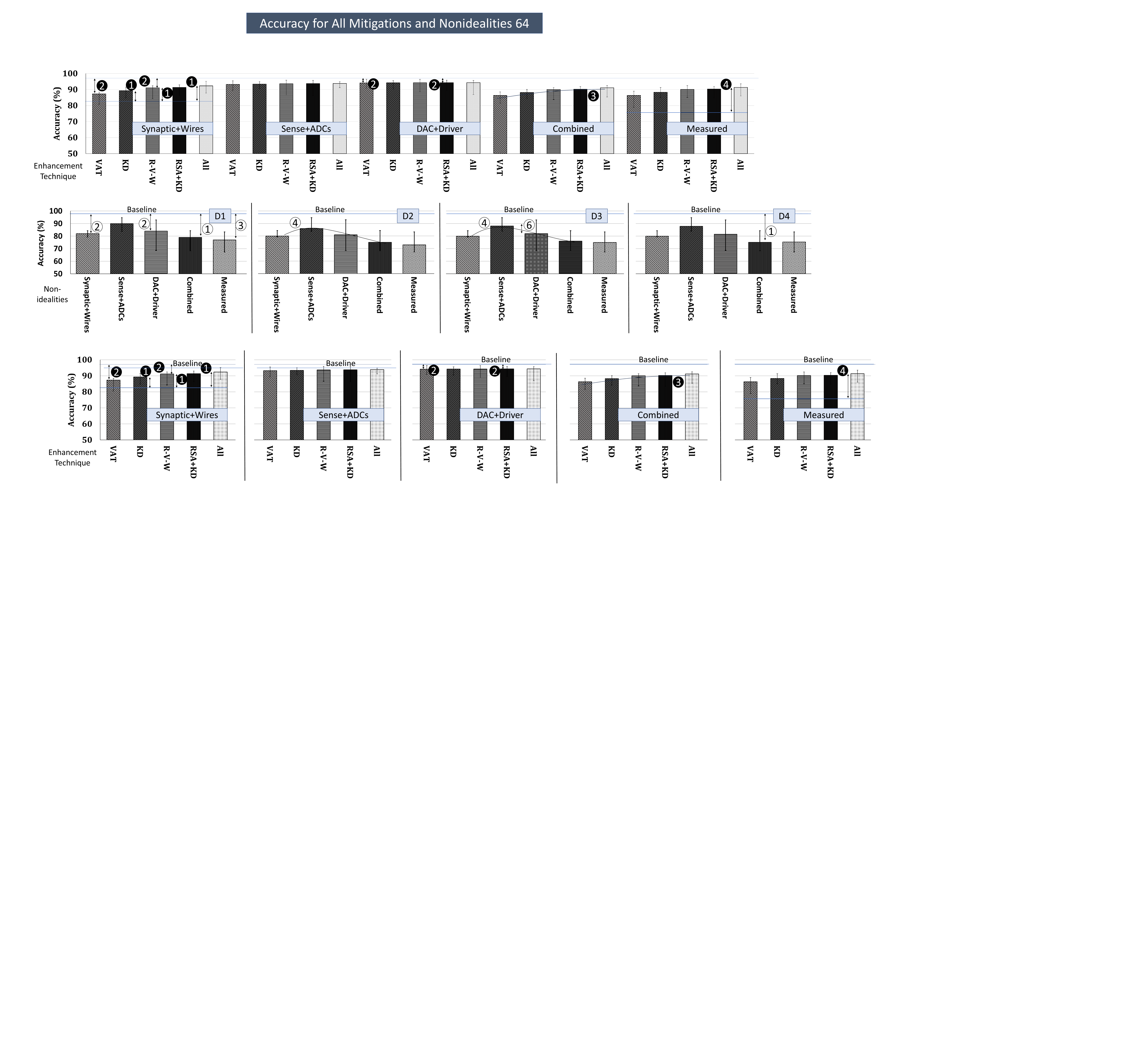}
        \caption{Accuracy after enhancement mechanisms for evaluated non-idealities on 64$\times$64 crossbars.}
        \label{fig:accuracy_after_allNonIdealities64_Bonito_LSTM_Architecture_With_Mitigation-evaluations_smart_mitigations}
    \end{minipage}\hfill
    \begin{minipage}{1\textwidth}
        \centering
        \includegraphics[width=1\textwidth]{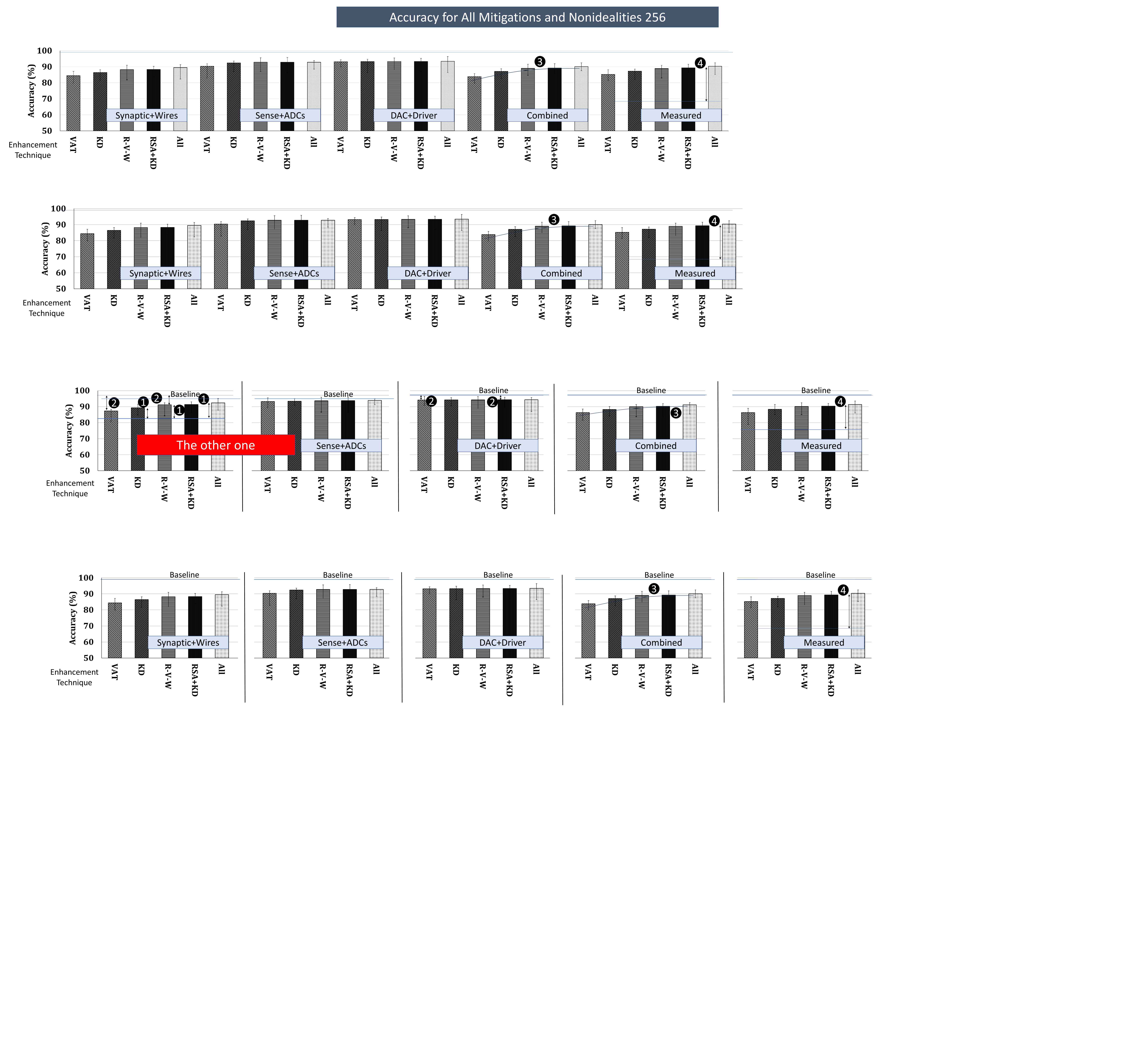} 
        \caption{Accuracy after enhancement mechanisms for evaluated non-idealities on 256$\times$256 crossbars.}
        \label{fig:accuracy_after_allNonIdealities256_Bonito_LSTM_Architecture_With_Mitigation-evaluations_smart_mitigations}
    \end{minipage}\hfill
    \label{fig:accuracy_after_allNonIdealities_Bonito_LSTM_Architecture_With_Mitigation-evaluations_smart_mitigations}
\end{figure*}

\subsubsection{Effect of Accuracy Enhancement on Write Variation.}
\label{subsec:accuracy_mitigations_for_write_variation-accuracy_mitigations_effects_nonidealities-evaluations_smart_mitigations}

\fig{\ref{fig:accuracy_after_writevariation_Bonito_LSTM_Architecture_With_Mitigation-evaluations_smart_mitigations}} presents the effects of our accuracy enhancement techniques (see \sect{\ref{subsec:accuracy_mitigator}}) considering different write variation rates across our four datasets (D1-D4). The horizontal dotted line shows the baseline accuracy using DFP 32-32 (see \sect{\ref{subsec:accuracy_quantization_effects-evaluations_notSmart}}) for the \bonito basecaller in all figures in \fig{\ref{fig:accuracy_after_writevariation_Bonito_LSTM_Architecture_With_Mitigation-evaluations_smart_mitigations}}.   \fig{\ref{fig:accuracy_after_writevariation_Bonito_LSTM_Architecture_With_Mitigation-evaluations_smart_mitigations}}-(a)-(d) evaluate the effect of \emph{\vat}, \emph{\kd}, \emph{\rvw}, \emph{\rsa{}+\kd} separately. \fig{\ref{fig:accuracy_after_writevariation_Bonito_LSTM_Architecture_With_Mitigation-evaluations_smart_mitigations}}-(e) considers all of our accuracy enhancement mechanisms together (\emph{Combined}), and \fig{\ref{fig:accuracy_after_writevariation_Bonito_LSTM_Architecture_With_Mitigation-evaluations_smart_mitigations}}-(f) averages the results of each accuracy enhancement technique over all the datasets (\emph{Averaged}).\footnote{The results in \fig{\ref{fig:accuracy_after_writevariation_Bonito_LSTM_Architecture_With_Mitigation-evaluations_smart_mitigations}} consider the cases in which \swordfish maps only \defaultPercentageMappedWeightOnChipMemoryRSAKD of weights to the \sram in our \rsa-based online retraining approach (see \sect{\ref{subsec:rsa_online_retraining-mitigation_techniques_for_accuracy}}). We will revisit this number in \sect{\ref{subsubsec:performance_and_throughput-evaluations_smart_mitigations}}.} We make four major observations from \fig{\ref{fig:accuracy_after_writevariation_Bonito_LSTM_Architecture_With_Mitigation-evaluations_smart_mitigations}}.

First, individual accuracy enhancement mechanisms evaluated in \fig{\ref{fig:accuracy_after_writevariation_Bonito_LSTM_Architecture_With_Mitigation-evaluations_smart_mitigations}}-(a)-(d) all improve the accuracy. However, their effectiveness reduces as the write variation rate increases.

Second, the online mechanism (\emph{\rsa{}+\kd}) in \fig{\ref{fig:accuracy_after_writevariation_Bonito_LSTM_Architecture_With_Mitigation-evaluations_smart_mitigations}}-(d) outperforms all the offline techniques in \fig{\ref{fig:accuracy_after_writevariation_Bonito_LSTM_Architecture_With_Mitigation-evaluations_smart_mitigations}}-(a)-(c). \emph{\rvw} in \fig{\ref{fig:accuracy_after_writevariation_Bonito_LSTM_Architecture_With_Mitigation-evaluations_smart_mitigations}}-(c)  comes second in terms of accuracy. However, the difference between \emph{\rsa{}+\kd} and \emph{\rvw} widens as the write variation rate increases.

Third, combining all the accuracy enhancement mechanisms (\emph{Combined} in \fig{\ref{fig:accuracy_after_writevariation_Bonito_LSTM_Architecture_With_Mitigation-evaluations_smart_mitigations}}-(e)) outperforms any individual technique over every single dataset and write variation rate.

Fourth, averaged over all the datasets (\emph{Averaged} in \fig{\ref{fig:accuracy_after_writevariation_Bonito_LSTM_Architecture_With_Mitigation-evaluations_smart_mitigations}}-(f)), \emph{Combined} mitigation techniques always produces the highest accuracy on average as well. However, on average, our online \emph{\rsa{}+\kd} technique achieves a close accuracy (less than \accuracyDiffRSAKDandCombinedBelowFifteenPercentWriteVariation difference) for low write variation rates, i.e., write variation less than 10\%.)

These results suggest that even with multiple accuracy enhancement techniques, only minor write variations (e.g., less than \maxMitigateWriteVariationAfterAllMitigations) can be tolerated. We conclude that a memristor-based \cim-enabeld accelerator for basecalling can be effective even with write variations, but such variations must be kept low (e.g., up to \maxMitigateWriteVariationAfterAllMitigations). Fortunately, the projected write variation rate for memristor-based devices~\cite{charan2020accurate-AccurateInferenceInaccurateMemory, jain2020rxnn-NonIdealities-ADCDACVariations} suggests the likelihood of achieving this percentage rate. For the rest of this manuscript, we assume a write variation of \maxMitigateWriteVariationAfterAllMitigations.

\subsubsection{Effect of Accuracy Enhancement for Combined Non-idealities.}
\label{subsec:accuracy_mitigations_for_combined_nonidealities-accuracy_mitigations_effects_nonidealities-evaluations_smart_mitigations}

\fig{\ref{fig:accuracy_after_allNonIdealities64_Bonito_LSTM_Architecture_With_Mitigation-evaluations_smart_mitigations}} presents the accuracy of basecalling with different accuracy enhancement techniques in crossbars of 64$\times$64 for the modeled non-idealities. For the non-idealities, we consider the five variations of \emph{Synaptic+Wires}, \emph{Sense+\adc}, \emph{\dac{}+Driver}, \emph{Combined}, and \emph{Measured} defined in \sect{\ref{subsec:combined_nonidealities-accuracy_nonidealities_effects-evaluations_notSmart}}. In \fig{\ref{fig:accuracy_after_allNonIdealities64_Bonito_LSTM_Architecture_With_Mitigation-evaluations_smart_mitigations}}, we evaluate five accuracy enhancement techniques of \emph{\vat}, \emph{\kd}, \emph{\rvw}, \emph{\rsa{}+\kd}, and \emph{All} (as defined in \sect{\ref{subsec:accuracy_mitigations_for_write_variation-accuracy_mitigations_effects_nonidealities-evaluations_smart_mitigations}}) per non-ideality.  \fig{\ref{fig:accuracy_after_allNonIdealities256_Bonito_LSTM_Architecture_With_Mitigation-evaluations_smart_mitigations}} presents the same experiments for crossbars of 256$\times$256. As we conclude in \sect{\ref{subsec:accuracy_mitigations_effects_nonidealities-evaluations_smart_mitigations}}, we assume \maxMitigateWriteVariationAfterAllMitigations write variation and \defaultPercentageMappedWeightOnChipMemoryRSAKD of the weights are mapped to the \sram in the online retraining approach (see \sect{\ref{subsec:rsa_online_retraining-mitigation_techniques_for_accuracy}}). We present our accuracy results averaged across all the evaluated datasets. We make four main observations from \fig{\ref{fig:accuracy_after_allNonIdealities64_Bonito_LSTM_Architecture_With_Mitigation-evaluations_smart_mitigations}}.

\begin{enumerate} [leftmargin=*]

    \item Combining of individual accuracy enhancement techniques does not improve the accuracy in an additive manner. For example, each of \emph{\vat}, \emph{\rvw}, and \emph{\rsa{}+\kd} in \fig{\ref{fig:accuracy_after_allNonIdealities64_Bonito_LSTM_Architecture_With_Mitigation-evaluations_smart_mitigations}}  improves accuracy due to \emph{Synaptic+Wires} by \flashbackAccuracyImprovementVATMitigationAveragedDatasetsSynapticSixtyfour, \flashbackAccuracyImprovementRVWMitigationAveragedDatasetsSynapticSixtyfour, \flashbackAccuracyImprovementRSAKDMitigationAveragedDatasetsSynapticSixtyfour, respectively. However, when we consider all non-idealities together in the \emph{All} configuration, accuracy improves by only \flashbackAccuracyImprovementALLMitigationAveragedDatasetsSynapticSixtyfour (\circled{1} in \fig{\ref{fig:accuracy_after_allNonIdealities64_Bonito_LSTM_Architecture_With_Mitigation-evaluations_smart_mitigations}}).

    \item The effectiveness of an individual accuracy enhancement technique depends on the underlying error and non-ideality it targets. For example, \emph{\vat} is as effective as \emph{\rsa{}+\kd} for non-idealities due to  \emph{\dac{}+Driver} (\accuracyImprovementVATMitigationAveragedDatasetsDACSixtyfour vs. \accuracyImprovementRSAKDMitigationAveragedDatasetsDACSixtyfour). However, the gap between the two approaches widens for non-idealities due to \emph{Synaptic+Wires}  (\accuracyImprovementVATMitigationAveragedDatasetsSynapticSixtyfour vs. \accuracyImprovementRSAKDMitigationAveragedDatasetsSynapticSixtyfour). See \circled{2} in \fig{\ref{fig:accuracy_after_allNonIdealities64_Bonito_LSTM_Architecture_With_Mitigation-evaluations_smart_mitigations}}.

    \item Accuracy enhancement techniques improve accuracy with a similar trend over different crossbar sizes (\circled{3} in \fig{\ref{fig:accuracy_after_allNonIdealities64_Bonito_LSTM_Architecture_With_Mitigation-evaluations_smart_mitigations}} and \fig{\ref{fig:accuracy_after_allNonIdealities256_Bonito_LSTM_Architecture_With_Mitigation-evaluations_smart_mitigations}}). Although these results are averaged over our datasets, one can make the same observation on each dataset as well.

    \item Accuracy enhancement techniques are more effective for larger crossbars than for smaller ones (e.g., 256$\times$256 compared to 64$\times$64). This is expected because there is more room for accuracy improvement for these larger crossbars, as their inaccuracies are higher. For example, we observe  \accuracyImprovementAllMitigationAveragedDatasetsMeasuredTwohundredfiftysix improvement in accuracy for 256$\times$256 crossbars (\circled{4} in \fig{\ref{fig:accuracy_after_allNonIdealities256_Bonito_LSTM_Architecture_With_Mitigation-evaluations_smart_mitigations}}) compared to \accuracyImprovementAllMitigationAveragedDatasetsMeasuredSixtyfour for 64$\times$64 (\circled{4} in \fig{\ref{fig:accuracy_after_allNonIdealities64_Bonito_LSTM_Architecture_With_Mitigation-evaluations_smart_mitigations}}), after all of the accuracy enhancement techniques are applied (\emph{All}) over all existing non-idealities (i.e., the \emph{Measured} configuration).

\end{enumerate}

We conclude that the basecalling accuracy of \accswordfish can match \sota levels by using robust techniques that build on each other employing reasonable crossbar sizes (e.g., 64$\times$64) and successfully accounting for substantial circuit variations, like write variations.

\subsection{Throughput Analysis of \accswordfish}
\label{subsubsec:performance_and_throughput-evaluations_smart_mitigations}

\begin{figure}[htbp]
\centering
    \includegraphics[width=1\linewidth]{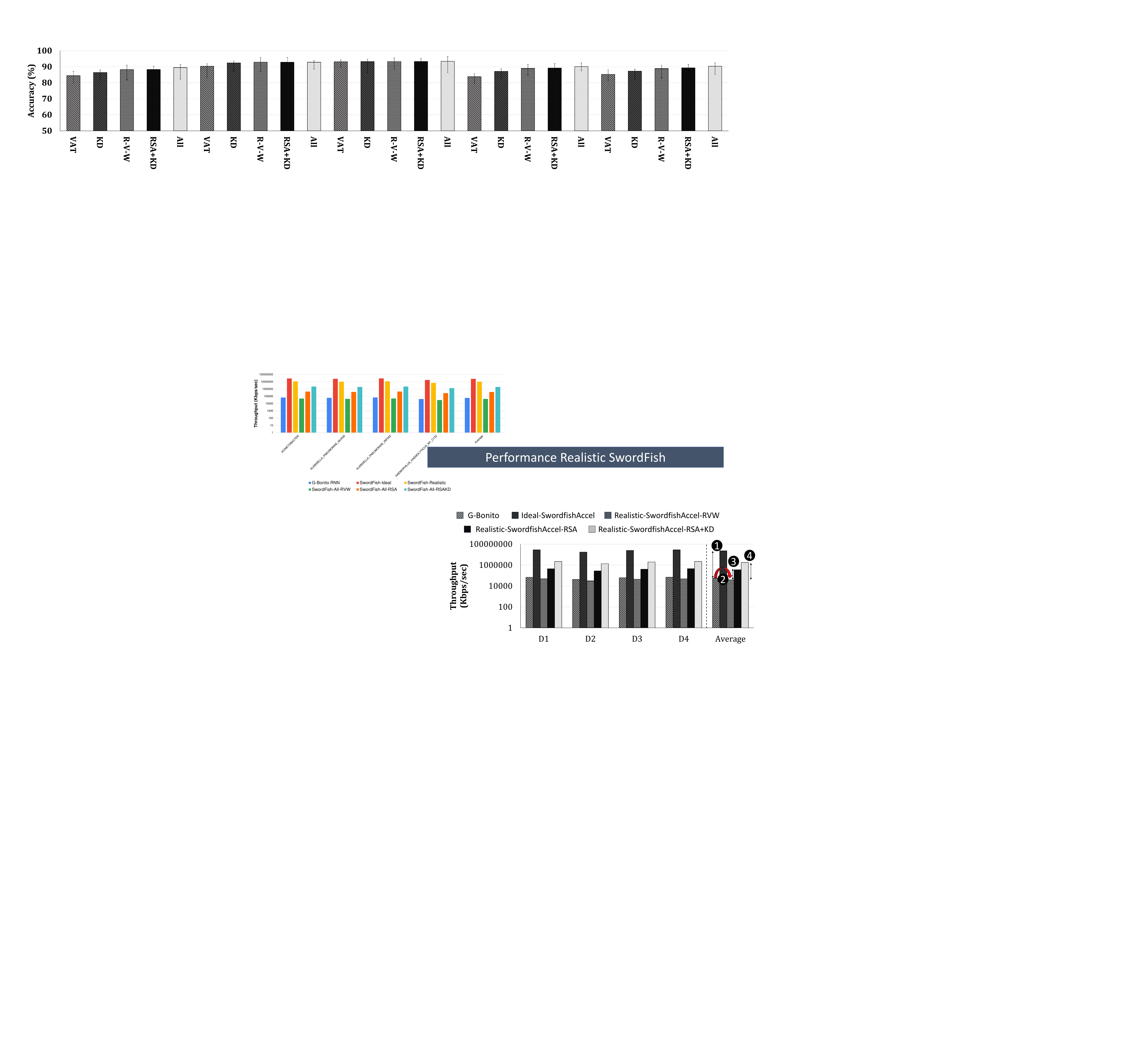}
    \caption{Throughput comparison of \swordfish variations.}
    \label{fig:performance_SwordFish_Bonito_LSTM-performance_and_energy-evaluations_smart_mitigations}
\end{figure}

\fig{\ref{fig:performance_SwordFish_Bonito_LSTM-performance_and_energy-evaluations_smart_mitigations}} shows the inference throughput for \bonito on a \gpu (\gbonito) card discussed in \sect{\ref{subsec:simulation_infrastructure-experimental_setup_and_methodology}}, \nomitigationhwaccswordfish,  \realhwaccswordfish-RVW, \realhwaccswordfish-\rsa, and \realhwaccswordfish-\rsa{}+\kd. We show the results for each of the four datasets and the average results over all datasets. The results are for a crossbar of size 64$\times$64 and a write variation rate of \maxMitigateWriteVariationAfterAllMitigations, and assuming \defaultPercentageMappedWeightOnChipMemoryRSAKD of weights are placed in \sram for Realistic-SwordfishAccel-RSA and \realhwaccswordfish-\rsa{}+\kd.

We make four key observations. First, \nomitigationhwaccswordfish improves the basecalling throughput over \gbonito{} for all datasets, by \throughputImprovementIdealAccswordfishVsBonitoAverage on average (\circled{1} in \fig{\ref{fig:performance_SwordFish_Bonito_LSTM-performance_and_energy-evaluations_smart_mitigations}}). We expect such a large improvement in throughput because \accswordfish is highly optimized for the main dominant kernel in the underlying \dnn of \bonito, namely \vmm, and avoids unnecessary data movement while harvesting the maximum parallelism.

Second, all versions of \realhwaccswordfish (i.e., Realistic-SwordfishAccel-RVW, \realhwaccswordfish-\rsa, and \realhwaccswordfish-\rsa{}+\kd) have lower performance than \nomitigationhwaccswordfish, irrespective of the dataset. Performance loss with a realistic \swordfish accelerator is expected because each realistic version adds overheads to mitigate accuracy loss due to realistically-modeled non-idealities, which directly affect the performance of a \vmm operation. For example, \rsa adds overheads due to (1)~the extra checks when reading some weights from the on-chip \sram memory and (2)~additional logic for combining the results from the memristor-based crossbar and on-chip memory readout.

Third, not all versions of \realhwaccswordfish outperform \gbonito. More specifically, if we use \rvw for mitigating non-idealities (\realhwaccswordfish-RVW in \fig{\ref{fig:performance_SwordFish_Bonito_LSTM-performance_and_energy-evaluations_smart_mitigations}}), the overhead due to additional verifications and writes significantly reduces the performance of basecalling throughput compared to \gbonito by \throughputDropRVWvsGbonitoAveragedDatasets on average (\circled{2} in \fig{\ref{fig:performance_SwordFish_Bonito_LSTM-performance_and_energy-evaluations_smart_mitigations}}).

Fourth, \realhwaccswordfish-\rsa and \realhwaccswordfish-\rsa{}+\kd provide, on average, \throughputImprovementAccRSAvsGbonitoAveragedDatasets and  \throughputImprovementAccRSAKDvsGbonitoAveragedDatasets higher throughput compared to \gbonito, respectively (\circled{3} and \circled{4} in \fig{\ref{fig:performance_SwordFish_Bonito_LSTM-performance_and_energy-evaluations_smart_mitigations}}). Note that, for the same accuracy, \realhwaccswordfish-\rsa{}+\kd\ requires fewer weights inside the \sram than \realhwaccswordfish-\rsa due to the retraining using \kd. Hence,  \realhwaccswordfish-\rsa{}+\kd is faster.

We conclude that a realistic basecalling accelerator designed using \swordfish by taking into account and mitigating all non-idealities of memristor-based \cim can significantly accelerate basecalling, yet its benefits are much lower than a corresponding accelerator that does not mitigate such non-idealities and thus has much lower accuracy.

\subsection{Area vs. Accuracy Analysis}
\label{subsubsec:area-evaluations_smart_mitigations}

\begin{figure}[htbp]
\vspace{0.5em}
\centering
    \includegraphics[width=1\linewidth]{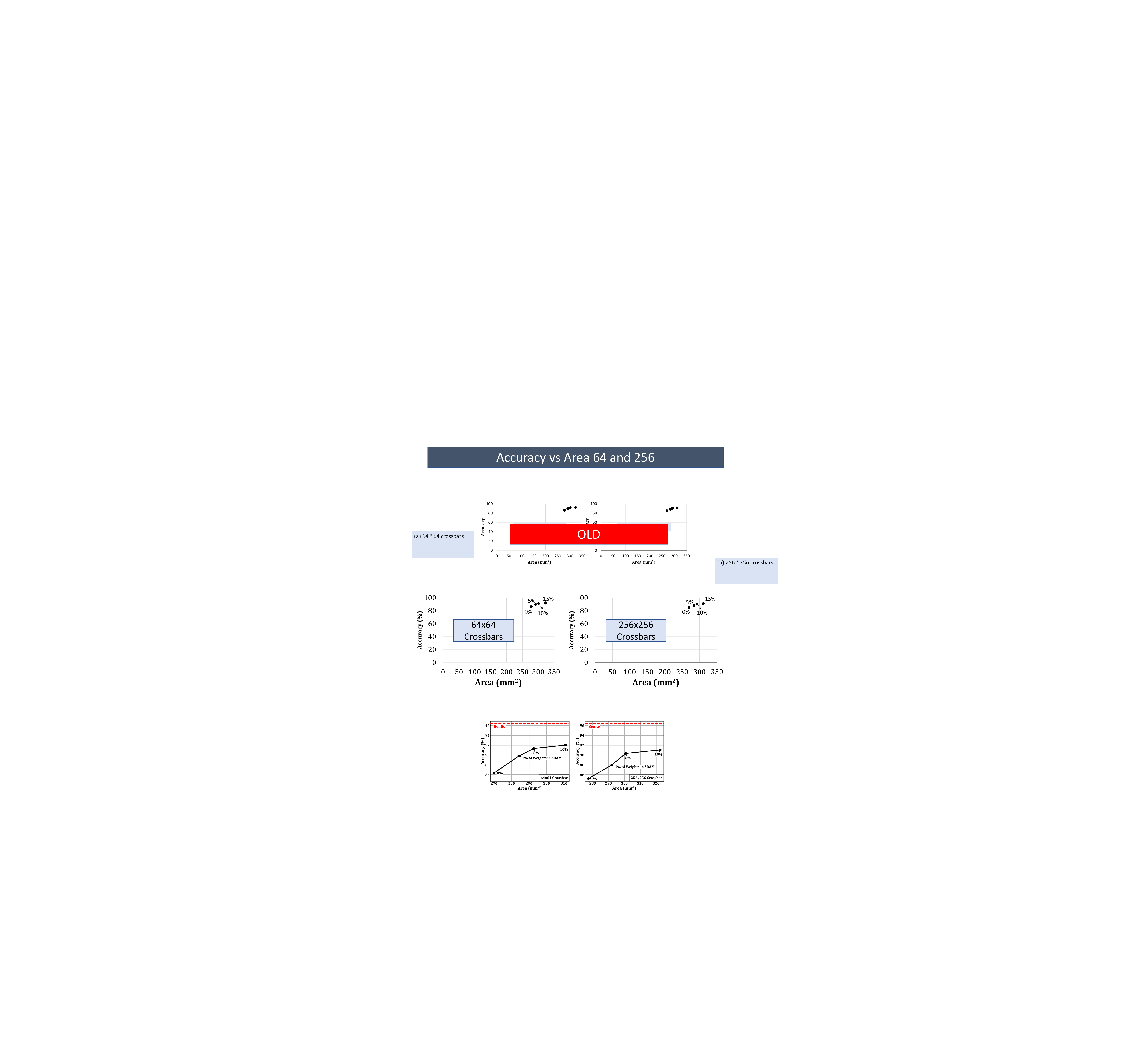}
    \caption{Accuracy vs. Area evaluation of \realhwaccswordfish-\rsa{}+\kd.}
    \label{fig:accuracy_vs_Area_RSAKD_64And256-evaluations_smart_mitigations}
\end{figure}

\fig{\ref{fig:accuracy_vs_Area_RSAKD_64And256-evaluations_smart_mitigations}} shows the tradeoff between accuracy and area in \realhwaccswordfish-\rsa{}+\kd (see \sect{\ref{subsubsec:performance_and_throughput-evaluations_smart_mitigations}}) for two different crossbar sizes (64$\times$64 on the left and 256$\times$256 on the right), with four different percentages of weights (i.e., 0\%, 1\%, 5\%, and 10\%) assigned to the \sram memory (see \sect{\ref{subsec:rsa_online_retraining-mitigation_techniques_for_accuracy}}). The area numbers show the absolute area for implementing \realhwaccswordfish-\rsa{}+\kd considering the overhead of \rsa{}+\kd discussed in \sect{\ref{subsec:rsa_online_retraining-mitigation_techniques_for_accuracy}}. The red dashed line shows the accuracy of the original \bonito basecaller. We make three main observations.

First, the more weights are assigned to \sram, the higher the accuracy of \realhwaccswordfish-\rsa{}+\kd. This is expected because we effectively reduce the non-idealities of the system by using more \sram cells to remap non-ideal memristors.

Second, the area of extra \sram cells used in \realhwaccswordfish-\rsa{}+\kd increases significantly with the percentage of weights assigned to \sram. In contrast, the accuracy improvement saturates and does not increase significantly beyond \defaultPercentageMappedWeightOnChipMemoryRSAKD of weights assigned to \sram.

Third, assigning only \defaultPercentageMappedWeightOnChipMemoryRSAKD of weights to \sram is sufficient to be within \accuracyRangeAfterRSAKD\ of \gbonito's accuracy for the 64$\times$64 crossbar.

We conclude that accounting for non-idealities in different ways exposes tradeoffs between accuracy and area overhead, which our \swordfish framework enables the designer to rigorously explore.

\section{Discussion and Future Work} \label{sec:discussions_limitations_futureWorks}

\subsection{Applicability of \swordfish Looking forward}

\swordfish emphasizes the importance of a framework for evaluating multiple metrics when designing a memristor-based \cim accelerator targeting large \dnn{}s that require throughput acceleration while having stringent bound for another metric, e.g., accuracy (in the presence of emerging technologies with many non-idealities).

\swordfish's realistic results, \realhwaccswordfish, for \bonito, a large \dnn, challenge the notion that \dnn-based applications naturally thrive on memristor-based \cim due to the inherent redundancy present in large neural networks. Although \realhwaccswordfish might not currently offer basecalling accuracy on par with \sotalong methods, its large (\throughputImprovementAccRSAKDvsGbonitoAveragedDatasets) enhancement in performance (\sect{\ref{subsubsec:performance_and_throughput-evaluations_smart_mitigations}}) at a much higher accuracy than baseline \cim marks it as an advantageous development. Even in the presence of memristor-based \cim non-ideality, \swordfish still shows promise, and  \realhwaccswordfish still maintains a competitive accuracy in basecalling by deploying a unique synergy of mitigation strategies (against non-idealities and variations) on moderately-large crossbar designs (e.g., 64$\times$64 or 256$\times$256). Our results in \sect{\ref{sec:evaluations_notSmart_and_smartMitigations}} detail this. Given our results, we believe it is productive and important to find more solutions to the memristor-based \cim non-idealities going forward; we believe some solutions will come with memristors becoming more mature, and some will come with more potent accuracy enhancement techniques and \hs methods.

\subsection{Other \dnn-based Applications}

Our paper discusses \swordfish as a framework for accelerating basecalling using a memristor-based \cim architecture. Our results (\sect{\ref{sec:evaluations_notSmart_and_smartMitigations}}) show the unique nature of the large \dnn in \bonito, which, despite its inherent redundancy, does not quite reach \sota accuracy on memristor-based \cim, thus presenting an exciting challenge. This intriguing finding encourages a deeper exploration into \cim designs for large \dnn{}s, reminding us not to rely solely on the scalability assumptions based on small network evaluations, such as simple \cnn{}s for \mnist. Our results also demonstrate a large acceleration opportunity for basecalling using \accswordfish if we can mitigate the memristor-induced accuracy loss through \hs approaches. We believe other \dnn-based applications that use memristor-based \cim accelerators (e.g., \cite{jain2020rxnn-NonIdealities-ADCDACVariations,  zhang2018deeproad, charan2020accurate-AccurateInferenceInaccurateMemory}) can also benefit from our approach and \swordfish. For example, large \dnn models in autonomous driving (e.g., \cite{zhang2018deeproad, kocic2019end, li2021testing}) that require accurate yet high-throughout and low-latency execution can use a \swordfish-like approach to build memristor-based \cim accelerators for their underlying large \dnn{}s. We believe and hope that \swordfish can aid such applications in terms of both accuracy and performance.

\subsection{Better Accuracy Enhancement Techniques}

Our results show that accuracy enhancement can pave the way toward \accswordfish becoming a reliable solution. Our online retraining mechanism shows the highest potential to improve the accuracy loss. We believe there needs to be more research on better mitigation techniques for existing and future non-idealities in memristor-based designs. Specifically, we suggest \hslong solutions such as our \rsa{}+\kd technique in \sect{\ref{subsec:rsa_online_retraining-mitigation_techniques_for_accuracy}}. Hardware-based solutions to mitigate non-idealities \cite{chen2017accelerator} that are orthogonal to our \rsa{}+\kd approach is also an example of possible avenues of future work.

\section{Related Work} \label{sec:relatedWork}

To our knowledge, \swordfish is the first framework that enables evaluating the acceleration of large \dnnlong{}s (\dnn{}s) on memristor-based \cimlong (\cim) designs considering hardware non-idealities. We have already compared \swordfish extensively to the currently-used version of the \bonito basecaller in \sect{\ref{sec:evaluations_notSmart_and_smartMitigations}} in terms of accuracy, throughput, and area overhead. This section briefly discusses related prior works on basecallers and \cim accelerators.

\subsection{Genomic Basecallers} \label{subsec:basecallers-relatedWork}

Several recent works propose approaches and techniques to either improve the accuracy of basecalling or accelerate it with minimum accuracy loss. These works take three main approaches: (1) new \dnn architectures (e.g.,~\cite{wick2019performance-Guppy, xu2021fast-FastBonito, singh2022framework-RUBICONQABASSkipClipRUBICALL-gaganbasecaller, dorado2022, flappie2018, Scrappie2019, bonito2020}), (2) new hardware platforms and designs such as GPUs and FPGAs to execute previously-proposed basecallers with minimum modifications (e.g.,~\cite{lou2020helix, singh2022framework-RUBICONQABASSkipClipRUBICALL-gaganbasecaller}), and (3) software techniques such as quantization to reduce the computation and storage overhead (e.g., \cite{esser2019learned-fixedPointQuantization1, jacob2018quantization-fixwedPointQuantization2, ding2017lightnn-reducedRepresentationPowerofTwoHWFriendlyQuantization3, li2019additive-HWFriendlyQuantization4, tambe2020algorithm-HWFriendlyQuantization5, singh2022framework-RUBICONQABASSkipClipRUBICALL-gaganbasecaller, xu2021fast-FastBonito}).

In contrast to these approaches, \swordfish is a framework for the \emph{evaluation} of \dnn-based (basecalling) accelerators. As such, \swordfish is orthogonal to prior works in basecalling, enabling proper evaluation of relevant works in the context of memristor-based in-memory acceleration.

\subsection{\cimlong Accelerators} 
\label{subsec:pimcim_Accs-relatedWork}

Many previous works investigate how to provide new functionality using compute-capable memories based on conventional (e.g., \cite{seshadri2017ambit, seshadri2015fast, ahn2016scalable, li2017drisa, aga2017computecache, ferreira2021pluto, gao2021parabit, park2022flash, gomez2022experimental}) and emerging memory technologies (e.g., \cite{lee2009architecting, xie2017scouting, strukov2008missing-HP, li2016pinatubo, zahedi2022system-MahdiVLSISOCSpecialSession, shafiee2016isaac, shahroodi2022demeter, shahroodi2022krakenonmem, ankit2019puma, singh2022cim-sttmramencryptionBNNCIMAbhairaj, prezioso2015training, chi2016prime, jeong2017parasitic-wireParasitics-SumitAnteneh, HDC-CIM-IBM}) to help solve the data movement overheads in today's systems. These works propose new functionality in at least three major categories: (1)~support for logical operations~(e.g.,~\cite{xie2017scouting, li2016pinatubo, seshadri2017ambit, zahedi2022system-MahdiVLSISOCSpecialSession, singh2022cim-sttmramencryptionBNNCIMAbhairaj, ni2017energy-VmminDNN2, cheng2019functional, strukov2008missing-HP}), (2)~support for complex operations, functions, and applications (e.g.,~\cite{shafiee2016isaac, aga2017computecache, shahroodi2022demeter, shahroodi2022krakenonmem, ferreira2021pluto, SparseMem_DATE2023_MahdiZahedi, li2017drisa, shahroodi2023Lightspeed-BNNOPCMLBRDate2023, oliveira2023transpimlib}), and (3)~programming and system support for the integration and adoption of such accelerators (e.g.,~\cite{ankit2019puma, chi2016prime, jain2020rxnn-NonIdealities-ADCDACVariations, zahedi2022system-MahdiVLSISOCSpecialSession, shafiee2016isaac, ahn2016scalable, Mahdi_arithmetic_MMM_twosComplement_IEEEAccess_2022, ahn2015pim, burr2017neuromorphic-VMMinDNN1, ni2017energy-VmminDNN2}).

Several prior works(e.g., \cite{jain2020rxnn-NonIdealities-ADCDACVariations, venkataramani2014axnn-DNNErrorResiliency2, charan2020accurate-AccurateInferenceInaccurateMemory, ahmetinci2022quidam}) investigate the new requirements, tradeoffs, and challenges that arise from using the \cim paradigm (e.g., dealing with non-idealities in the analog operations). To our knowledge, no work has proposed a complete solution or framework for these challenges; thus, this area requires further investigation.

\swordfish aligns with these works as it provides (1) new functionality for compute-capable memristors at the application level for accelerating genomic basecalling and (2) a framework for evaluating the practical challenges posed by the non-idealities in the memristor computation through mitigation techniques.

\section{Conclusion} \label{sec:conclusion}

This paper introduces \swordfish, a modular and extensible framework for accelerating the evaluation of genomic basecalling via a memristor-based \cimlong architecture. \swordfish includes a strong evaluation methodology, mitigation strategies for hardware non-idealities, and characterization results to guide the modeling of memristors. Using \swordfish, we demonstrate the significant challenges of using non-ideal memristor-based computations for genomic basecalling and how to solve them by combining multiple mitigation techniques at the circuit and system levels. We demonstrate the usefulness of our findings by developing \accswordfish, a concrete memristor-based \cim design for our target basecaller \bonito that uses accuracy enhancement techniques guided by \swordfish. We conclude that the \swordfish framework effectively facilitates the development and adoption of memristor-based CIM designs for basecalling, which we hope will be leveraged by future work. We also believe that our framework is applicable to other \dnn-based applications and hope future work takes advantage of this.

\begin{acks}
  {
    We thank the anonymous reviewers of MICRO 2023 for their valuable feedback. We thank the members of the QCE department at TU Delft and the SAFARI Research Group at ETH Zurich for valuable feedback and the stimulating intellectual environment they provide. We acknowledge the generous gifts provided by our industrial partners, including Google, Huawei, Intel, Microsoft, and VMware. This research was partially supported by the EU Horizon project BioPIM (grant agreement 101047160), the AI Chip Center for Emerging Smart Systems Limited (ACCESS), the Swiss National Science Foundation (SNSF),  Semiconductor Research Corporation (SRC), and the ETH Future Computing Laboratory (EFCL).
  }
\end{acks}




\bibliographystyle{ACM-Reference-Format}
\bibliography{main}

\end{document}